\documentclass[10pt, a4paper]{article}
\usepackage{amsmath}
\usepackage{amssymb}
\usepackage[]{graphicx}
\usepackage[pdfauthor={P. P. Cook and P. C. West},
pdftitle={$M$-Theory Solutions in Multiple Signatures from $E_{11}$}]{hyperref}
\begin{document}
\title{$M$-Theory Solutions in Multiple Signatures from $E_{11}$}
\author{Paul P. Cook\footnote{email: paul.p.cook@kcl.ac.uk} and Peter C. West\footnote{email: pwest@mth.kcl.ac.uk}\\ \and \\{\itshape Department of Mathematics, King's College London,\/}\\{\itshape Strand, London WC2R 2LS, U.K.\/}}
\begin{titlepage}
\begin{flushright}
Preprint KCL-MTH-05-06\\
{\tt hep-th/0506122}
\end{flushright}
\vspace{70pt}
\centering{\LARGE $M$-Theory Solutions in Multiple Signatures from $E_{11}$}\\
\vspace{30pt}
\def\thefootnote{\fnsymbol{footnote}}
Paul P. Cook\footnote{\href{mailto:paul.p.cook@kcl.ac.uk}{email: paul.p.cook@kcl.ac.uk}} and Peter C. West\footnote{\href{mailto:pwest@mth.kcl.ac.uk}{email: pwest@mth.kcl.ac.uk}}\\
\setcounter{footnote}{0}
\vspace{10pt}
{\itshape Department of Mathematics, King's College London,\/}\\
{\itshape Strand, London WC2R 2LS, U.K.\/}\\
\vspace{10pt}
\vspace{30pt}
\begin{abstract}
We generalise the previously given $E_{11}$ half BPS solution generating group element to general weights of $A_{10}$. We find that it leads to solutions of $M$-theory but in signatures $(1,10)$, $(2,9)$, $(5,6)$, $(6,5)$, $(9,2)$ and $(10,1)$. The signature transformations of the solution are naturally generated by the Weyl reflections required to transform the lowest $A_{10}$ weight into a general weight in the same representation. We also rediscover known $S$-brane solutions in $M$-theory from the group element in different signatures.
\end{abstract}
\end{titlepage}
\clearpage
\newpage
\section{Introduction}
It has been conjectured that $M$-theory possesses a large Kac-Moody symmetry which is non-linearly realised and thought to contain $E_{11}$\cite{West1}. Indeed, it was shown that the bosonic sector of eleven dimensional supergravity could be described by a non-linear realisation \cite{West2}, that is, if one assumes that the non-linear realisation was extended such that it was a Kac-Moody algebra then this algebra must contain $E_{11}$. Subsequently, two variants of this conjecture have been made \cite{DamourHenneauxNicolai, EnglertHouart}. The first differs from the original suggestion of \cite{West1} in that it only adopted the sub-algebra $E_{10}$ as a symmetry. However, it also differed from the way the original suggestion of \cite{West1}, subsequently evolved in \cite{West3}, where spacetime was incorporated by extending the translation generators to be part of an $E_{11}$ representation. Indeed, in \cite{DamourHenneauxNicolai} spacetime was supposed to be contained in effect in the Kac-Moody algebra. The proposal of \cite{EnglertHouart} also adopted $E_{11}$ as the symmetry but took a similar approach to spacetime as that of \cite{DamourHenneauxNicolai}. For a discussion of the merits of these approaches to spacetime see reference \cite{KleinschmidtWest}. 

One of the hopes for such a large symmetry algebra is that it might provide insights into the solutions of the theory. Indeed, relationships have been uncovered between the half BPS solution of eleven dimensional supergravity and $E_{11}$ and more generally between all the oxidised theories and the suspected $G^{+++}$ symmetries of their extensions \cite{EnglertHouart,West}. In \cite{West, CookWest} it was pointed out that any solution of the non-linearly realised $E_{11}$ theory could be expressed as an $E_{11}$ group element and it was  found that all the known half BPS solutions were generated by a very simple group element which was specified by the choice of a single root of $E_{11}$,
\begin{equation}
g=\exp(-{\frac{1}{(\beta,\beta)}\ln N }\beta \cdot H)\exp((1-N)E_\beta) \label{groupelement}
\end{equation}
The root, $\beta$, is $\alpha_1+2\alpha_2+3(\alpha_3+\ldots \alpha_8)+2\alpha_9+\alpha_{10}+\alpha_{11}$ for the $M2$-brane and $\alpha_1+2\alpha_2+3\alpha_3+4\alpha_4+5\alpha_5+6\alpha_6+5\alpha_7+4\alpha_8+3\alpha_9+2\alpha_{10}+2\alpha_{11}$ for the $M5$-brane. In fact \cite{West} only used a subset of the possible roots of $E_{11}$ and in particular those which when decomposed to the $A_{10}$ sub-algebra of $E_{11}$ corresponded to lowest weights of $A_{10}$. Indeed one can explicitly check that the $M2$ and $M5$ branes are electric solutions for the lowest weights, see appendix \ref{electricbranesolutions}.

In this paper we will consider the other weights of $E_{11}$. We will find that the group element does indeed lead to solutions for all weights. However, some of these are to $M$-theories which are not in the $(1,10)$ signature. Since the different weights are related by Weyl transformations these will play an important role in this paper. The Weyl transformations of $E_{11}$ were found to correspond in the dimensionally reduced theory to the U-duality transformations \cite{EnglertHouartTaorminaWest}. Weyl transformations of some solutions in eleven dimensions were discussed in \cite{West3,EnglertHouart,EnglertHouartTaorminaWest,EnglertHenneauxHouart}. 

The non-linearly realised theory requires that the local sub-algebra is specified. This was initially taken to be the Cartan invariant involution, but this lead to a Euclidean theory and to find the usual theory in $(1,10)$ signature required a simple Wick rotation. However, it was realised that one could directly find the theory in $(1,10)$ signature by adopting  a different local sub-algebra by inserting a single minus sign in the generating set of the local sub-algebra as compared to that for the Cartan invariant involution. This was intially suggested in \cite{West3} (see equations 2.17-19) and a more systematic discussion was given in reference \cite{EnglertHouart} and subsequently in \cite{KleinschmidtWest}. It was then realised \cite{Keurentjes,West4} that by inserting other minus signs that one could find $M$-theories in other signatures. Furthermore, it was shown by Keurentjes \cite{Keurentjes}, that the Weyl reflections of $E_{11}$ do not commute with the choice of local sub-algebra, with the consequence that, by carrying out Weyl transformations, the usual signature of $M$-theory, $(1,10)$, occurs within $E_{11}$ alongside the signatures $(2,9)$, $(5,6)$ and their inverses. These were the theories that were called the $M*$ and $M'$-theories, respectively, in reference \cite{Hull}. The relation between $E_{11}$ and $E_{10}$ was discussed in \cite{EnglertHenneauxHouart} and the relationship between the different 10-dimensional theories and Weyl reflections was elaborated. Recently, this line of thought has been extended to all $G^{++}$ theories \cite{deBuylHouartTabti}.

In section two we evaluate the bosonic equations of motion of eleven dimensional supergravity in an arbitrary signature for a brane ansatz that includes the possibility of several times in the world-volume of the brane and also in the transverse space. We show how given one solution to a theory in a given signature we can generate classes of solutions in theories of different signature by inverting the signature in the transverse space of the solution, and in certain cases by inversion of the brane world-volume signature. These considerations are extended to find all possible signatures for which there exist solutions related to the $M2$ and $M5$-branes.

In section three we first review the work of Keurentjes on the effect of $E_{11}$ Weyl transformations on the signature of the theory. We then examine the effect of the Weyl transformation on the well known $M2$, $M5$ and $pp$-wave solutions in the $(1,10)$ theory. Commencing with the group element (\ref{groupelement}) and the lowest weights of $A_{10}$, different weights are obtained by Weyl reflections. Since it is possible to use a series of Weyl reflections that alter the signature but not the root associated to the weight we find that each weight, including the lowest, has an ambiguity concerning which signature it exists in. Indeed, we find all possible such transformed solutions and the corresponding signatures of the theory in
which each solution should exist. We compare these with the results of section two and find that the Weyl transformations do indeed lead to solutions in the theories of the required signature, exactly reproducing all the known solutions of the $M*$ and $M'$-theories \cite{Hull}. However, the Weyl transformations also lead to examples which are not solutions in the theory of the given signature. In these cases, which are as numerous as the solutions, the example is a solution to a theory with an alternative sign in front of the kinetic $F^2$ term in the action.

Spacelike brane solutions, or $S$-branes, are branes which map out a spacelike world-volume \cite{GutperleStrominger}.\footnote{The convention for naming spacelike branes is that an S$p$-brane has a $(p+1)$ Euclidean world-volume.} Consequently an $S$-brane only exists for a moment in time. We show that these are naturally contained in the group element (\ref{groupelement}). In section \ref{spacelikeinvolution}, we propose that a choice of local sub-algebra which invokes a time coordinate in the transverse space of an $M$-theory p-brane solution, as oppose to inducing the usual longitudinal time coordinate, reproduces the known $S2$ and $S5$-branes of $M$-theory \cite{ChenGaltsovGutperle}. We also find encoded in the group element (\ref{groupelement}) their relations in $M*$ and $M'$-theories.

\section{General Signature Formulation of the Einstein Equations} \label{general_signature_einstein_equations}
In this paper we will be working beyond the usual signatures of supergravity, it will be useful to express the Einstein equations in a form that is applicable to different signatures. The Einstein equations and the gauge equations for a single brane solution can be derived by varying the truncated form of a gravity action, where the truncation is the restriction to the kinetic term for one of the theory's n-form field strengths, the dilaton term and the Ricci scalar, e.g.
\begin{equation}
A=\frac{1}{16\pi G_{D}}\int {d^{D}x\sqrt{-g}(R-\frac{1}{2}\partial^\mu\phi\partial_\mu\phi-\frac{1}{2.n!}e^{a_i\phi}F_{\mu_1\ldots \mu_n}F^{\mu_1\ldots \mu_n})} \label{action}
\end{equation}
The usual Chern-Simons term appearing in the eleven dimensional supergravity action has been omitted here since for the class of extremal branes we will consider it plays no role in the dynamics, and will not affect our discussions. The equations of motion determined by varying with respect to the metric, $g_{\mu\nu}$, are the Einstein equations and the equation that comes from varying the gauge field is the gauge equation. There is also an equation of motion coming from the variation of the dilaton field. For the generic truncated action these are
\begin{align}
\nonumber &{R^\mu}_\nu \nonumber = \frac{1}{2}\partial^\mu\phi\partial_\nu\phi+\frac{1}{2n!}e^{a_i\phi}(nF^{\mu \lambda_2\ldots \lambda_{n}}F_{\nu \lambda_2\ldots \lambda_{n}} -\frac{n-1}{D-2}{\delta^\mu}_\nu F_{\lambda_1\ldots \lambda_{n}}F^{\lambda_1\ldots \lambda_{n}})\\
&\partial_\mu(\sqrt{-g}e^{a_i\phi}F^{\mu \lambda_2 \ldots \lambda_{n}})= 0 \label{EinsteinEq}\\
\nonumber &\frac{1}{\sqrt{-g}}\partial_\mu(\sqrt{-g}\partial^\mu\phi)-\frac{a_i}{2.n!}e^{a_i\phi}F_{\mu \lambda_2 \ldots \lambda_{n}}F^{\mu \lambda_2 \ldots \lambda_{n}}= 0
\end{align}
We compute the curvature components in the spin-connection formalism as described in \cite{Argurio}, but we commence with a line element for a more general than usual brane solution,
\begin{equation}
ds^2=A^2(\sum_{i=1}^{i=q}-dt_{i}^2+\sum_{j=1}^{j=p}dx_{j}^2)+B^2(\sum_{a=1}^{a=c}-du_{a}^2+\sum_{b=1}^{b=d}dy_{b}^2) \label{ansatz}
\end{equation}
The coordinates are split into two groups, those that are longitudinal to the brane, $t_i$ and $x_i$, we indicate with indices $\{i,j,k, \ldots \}$, and those that are transverse, $u_a$ and $y_a$ with $\{a,b,c, \ldots \}$. The given line element is the world-volume of a brane with signature $(q,p)$ on the brane and signature $(c,d)$ in the bulk; the corresponding global signature is $(q+c,p+d)$. We adopt the notation $[(q,p),(c,d)]$ to express a single signature for our ansatz in terms of its longitudinal, $(q,p)$ and transverse components $(c,d)$. For a global signature $(t,s)$ then $q+c=t$ and $p+d=s$. The case when $c=0$ and $q=1$, corresponds to the usual single brane solutions. In eleven-dimensional supergravity there is no dilaton, if we set the dilaton coupling to zero, the coefficients $A$ and $B$, functions of the transverse coordinates $(u_a,y_a)$, take the form in the extremal case \cite{Argurio, Stelle, Ohta1} 
\begin{equation}
A=N_{(c,d)}^{-(\frac{1}{p+1})},\qquad B=N_{(c,d)}^{(\frac{1}{D-p-3})} \label{singlebranecoefficients}
\end{equation}
Where, 
\begin{equation}
N_{(c,d)}=1+\frac{1}{D-p-3}\sqrt{\frac{\Delta}{2(D-2)}}\frac{\|\bf Q\|}{r^{(D-p-3)}} \label{harmonicfunction}
\end{equation}
Where $\bf Q$ is the conserved charge associated with the $p$-brane solution, $\Delta=(p+1)(d-2)+\frac{1}{2}a_i^2(D-2)$ and $r$ is the radial distance in the transverse coordinates such that $r^2=-u_au_a+y_by_b$. That is, $N_{(c,d)}$ are independent of the longitudinal coordinates, and are harmonic functions in the transverse coordinates, $u_a$ and $y_b$, so that $\partial^\mu\partial_\mu N_{(c,d)}=0$.

The full non-zero curvature components are,
\begin{align}
\nonumber {R^{t_i}}_{t_i}=B^{-2} \{ &\partial_{u_a}\partial_{u_a}\ln{A}{\hat{\delta}^{u_a}}_{\hspace{7pt} u_a}-\partial_{y_a}\partial_{y_a}\ln{A}{\hat{\delta}^{y_a}}_{\hspace{7pt} y_a}\\
\nonumber &+\partial_{u_a}\ln{A}\partial_{u_a}\Psi{\hat{\delta}^{u_a}}_{\hspace{7pt} u_a}-\partial_{y_a}\ln{A}\partial_{y_a}\Psi{\hat{\delta}^{y_a}}_{\hspace{7pt} y_a} \} {\hat{\delta}^{t_i}}_{\hspace{7pt} t_i} \\
\nonumber {R^{x_i}}_{x_i}=B^{-2} \{ &\partial_{u_a}\partial_{u_a}\ln{A}{\hat{\delta}^{u_a}}_{\hspace{7pt} u_a}-\partial_{y_a}\partial_{y_a}\ln{A}{\hat{\delta}^{y_a}}_{\hspace{7pt} y_a}\\
\nonumber&+\partial_{u_a}\ln{A}\partial_{u_a}\Psi{\hat{\delta}^{u_a}}_{\hspace{7pt} u_a}-\partial_{y_a}\ln{A}\partial_{y_a}\Psi{\hat{\delta}^{y_a}}_{\hspace{7pt} y_a} \} {\hat{\delta}^{x_i}}_{\hspace{7pt} x_i} \\
\nonumber {R^{u_a}}_{u_a}=B^{-2} \{&\partial_{u_a}\partial_{u_a}\ln{B}{\hat{\delta}^{u_a}}_{\hspace{7pt} u_a}-\partial_{y_a}\partial_{y_a}\ln{B}{\hat{\delta}^{y_a}}_{\hspace{7pt} y_a}\\
 &+\partial_{u_a}\ln{B}\partial_{u_a}\ln{B}({\hat{\delta}^{y_a}}_{\hspace{7pt}y_a}+{\hat{\delta}^{u_a}}_{\hspace{7pt} u_a}-2)\\
\nonumber &+\partial_{u_a}\ln{A}\partial_{u_a}\ln{A}({\hat{\delta}^{t_i}}_{\hspace{7pt}t_i}+{\hat{\delta}^{x_i}}_{\hspace{7pt} x_i})
-\partial_{y_a}\ln{B}\partial_{y_a}\Psi{\hat{\delta}^{y_a}}_{\hspace{7pt} y_a}\\
\nonumber &+\partial_{u_a}\partial_{u_a}\Psi+\partial_{u_a}\ln{B}\partial_{u_a}\Psi({\hat{\delta}^{u_a}}_{\hspace{7pt} u_a}-2)\}{\hat{\delta}^{u_a}}_{\hspace{7pt} u_a}\\
\nonumber {R^{y_a}}_{y_a}=B^{-2} \{&\partial_{u_a}\partial_{u_a}\ln{B}{\hat{\delta}^{u_a}}_{\hspace{7pt} u_a}-\partial_{y_a}\partial_{y_a}\ln{B}{\hat{\delta}^{y_a}}_{\hspace{7pt} y_a}\\
\nonumber &-\partial_{y_a}\ln{B}\partial_{y_a}\ln{B}({\hat{\delta}^{y_a}}_{\hspace{7pt}y_a}+{\hat{\delta}^{u_a}}_{\hspace{7pt} u_a}-2)\\
\nonumber &-\partial_{y_a}\ln{A}\partial_{y_a}\ln{A}({\hat{\delta}^{t_i}}_{\hspace{7pt}t_i}+{\hat{\delta}^{x_i}}_{\hspace{7pt} x_i})
+\partial_{u_a}\ln{B}\partial_{u_a}\Psi{\hat{\delta}^{u_a}}_{\hspace{7pt} u_a}\\
\nonumber &-\partial_{y_a}\partial_{y_a}\Psi-\partial_{y_a}\ln{B}\partial_{y_a}\Psi({\hat{\delta}^{y_a}}_{\hspace{7pt} y_a}-2)\}{\hat{\delta}^{y_a}}_{\hspace{7pt} y_a}
\end{align}
Where ${\hat{\delta}^{x_i}}_{\hspace{7pt} x_i}$ counts the number of $x_i$ coordinates in the line element (e.g. for the $M2$-brane, ${\hat{\delta}^{t_i}}_{\hspace{7pt} t_i}=1$, ${\hat{\delta}^{x_i}}_{\hspace{7pt} x_i}=2$, ${\hat{\delta}^{u_a}}_{\hspace{7pt} u_a}=0$ and ${\hat{\delta}^{y_a}}_{\hspace{7pt} y_a}=8$); repeated lowered indices a, b, i and j are not summed - sums are taken care of via the counting symbols $\hat{\delta}$; and, for the ansatz (\ref{ansatz}) with extremal coefficients A and B (\ref{singlebranecoefficients}),
\begin{align}
\nonumber \Psi\equiv &(p+1)\ln{A}+(D-p-3)\ln{B}\\
=&(p+1)\ln{N_p^{-2(\frac{1}{p+1})}}+(D-p-3)\ln{N_p^{2(\frac{1}{D-p-3})}}\\
\nonumber =&0
\end{align}
The curvature terms reduce to
\begin{align}
\nonumber {R^{t_i}}_{t_i}=B^{-2} \{ &\partial_{u_a}\partial_{u_a}\ln{A}{\hat{\delta}^{u_a}}_{\hspace{7pt} u_a}-\partial_{y_a}\partial_{y_a}\ln{A}{\hat{\delta}^{y_a}}_{\hspace{7pt} y_a} \} {\hat{\delta}^{t_i}}_{\hspace{7pt} t_i} \\
\nonumber {R^{x_i}}_{x_i}=B^{-2} \{ &\partial_{u_a}\partial_{u_a}\ln{A}{\hat{\delta}^{u_a}}_{\hspace{7pt} u_a}-\partial_{y_a}\partial_{y_a}\ln{A}{\hat{\delta}^{y_a}}_{\hspace{7pt} y_a} \} {\hat{\delta}^{x_i}}_{\hspace{7pt} x_i} \\
\nonumber {R^{u_a}}_{u_a}=B^{-2} \{&\partial_{u_a}\partial_{u_a}\ln{B}{\hat{\delta}^{u_a}}_{\hspace{7pt} u_a}-\partial_{y_a}\partial_{y_a}\ln{B}{\hat{\delta}^{y_a}}_{\hspace{7pt} y_a}\\
&+\partial_{u_a}\ln{B}\partial_{u_a}\ln{B}({\hat{\delta}^{y_a}}_{\hspace{7pt}y_a}+{\hat{\delta}^{u_a}}_{\hspace{7pt} u_a}-2) \label{curvatureterms}\\
\nonumber &+\partial_{u_a}\ln{A}\partial_{u_a}\ln{A}({\hat{\delta}^{t_i}}_{\hspace{7pt}t_i}+{\hat{\delta}^{x_i}}_{\hspace{7pt} x_i})\}{\hat{\delta}^{u_a}}_{\hspace{7pt} u_a}\\
\nonumber {R^{y_a}}_{y_a}=B^{-2} \{&\partial_{u_a}\partial_{u_a}\ln{B}{\hat{\delta}^{u_a}}_{\hspace{7pt} u_a}-\partial_{y_a}\partial_{y_a}\ln{B}{\hat{\delta}^{y_a}}_{\hspace{7pt} y_a}\\
\nonumber &-\partial_{y_a}\ln{B}\partial_{y_a}\ln{B}({\hat{\delta}^{y_a}}_{\hspace{7pt}y_a}+{\hat{\delta}^{u_a}}_{\hspace{7pt} u_a}-2)\\
\nonumber &-\partial_{y_a}\ln{A}\partial_{y_a}\ln{A}({\hat{\delta}^{t_i}}_{\hspace{7pt}t_i}+{\hat{\delta}^{x_i}}_{\hspace{7pt} x_i})
\}{\hat{\delta}^{y_a}}_{\hspace{7pt} y_a}
\end{align}
The $M2$, $M5$, and $pp$-wave solutions \cite{BPSsolutions} of $M$-theory are encoded in a group element of the non-linear realisation of $E_{11}$ \cite{West}. For reference and comparison with later solutions, we demonstrate that these cases satisfy the Einstein equations in appendix \ref{electricbranesolutions}.

\subsection{New Solutions from Signature Change} \label{solution_preserving}
A change of signature has the potential to alter both the Einstein equations and the field content of a theory. In preparation for the next section, we pose a question: given a solution to the Einstein equations in one signature, are there any other signatures which would also carry a related version of that solution in the new signature? We note that the harmonic function, $N_{(c,d)}$, is a function of the transverse coordinates and may be transformed by a signature change. By a 'related solution' we specifically mean that if the Einstein equations for a given solution were reexpressed in terms of the functions carrying the new signature then they would remain balanced and we would find a new solution.

Signature change can be brought about in two equivalent ways, the first is as a mapping of a coordinate, or a subset of the coordinates, $x^\mu\rightarrow ix^\mu, x_\mu\rightarrow -i x_\mu$, leaving the metric unaltered. For example, a Lorentzian signature can be made Euclidean by making the change on the temporal coordinate, $t^1\rightarrow ix^1$, having the effects,
\begin{align}
\nonumber ds^2 &= g_{\mu\nu}x^\mu x^\nu \\
\nonumber &=g_{t_1t_1}dt^2_1+g_{x_2x_2}dx^2_2+\ldots +g_{x_Dx_D}dx^2_D \\
\nonumber &=-f_1(N)dt^2_1+f_2(N)dx^2_2+\ldots +f_D(N)dx^2_D \\
& \downarrow \label{euclideanisation}\\
\nonumber ds^2&=f_1(N)dx^2_1+f_2(N)dx^2_2+\ldots +f_D(N)dx^2_D \\
&\nonumber = g_{x_1x_1}dx^2_1+g_{x_2x_2}dx^2_2+\ldots +g_{x_Dx_D}dx^2_D
\end{align}
Where $\pm f_i(N)$, some function of $N$, is the metric component in each case. An electric field $A_\mu$ transforms as
\begin{equation}
A_\mu dx^\mu = A_{t_1}dt^1 \rightarrow iA_{x_1}dx^1 \label{Wickrotation}
\end{equation}
Equivalently, signature change at the quadratic level of the line element can be thought of as a transformation of the metric components, or subset of the metric components, where appropriate, $g_{\mu\nu}\rightarrow -g_{\mu\nu}$ as opposed to the coordinates. Correspondingly we can view the example above (\ref{euclideanisation}), as the transformation $g_{t_1t_1}\rightarrow -g_{x_1x_1}$. Both methods realise the change of line element but only when applied independently.

In equation (\ref{curvatureterms}) we have written out the curvature coefficients for our ansatz, (\ref{ansatz}). We observe that the expressions for ${R^{t_i}}_{t_i}$ and ${R^{x_i}}_{x_i}$ satisfying our explicit ansatz of (\ref{ansatz}) are interchanged under the interchange of longitudinal temporal and spatial coordinates, given by $t^i\rightarrow ix^i$ and $x^j\rightarrow it^j$. These transformations corresponds to the notational swap $t_i\leftrightarrow x_i$ in the equations (\ref{curvatureterms}) and the set of curvature terms as a whole is unaffected. In terms of signature this corresponds to a signature inversion on only the longitudinal coordinates. Alternatively, the swap $u_a$ for $y_a$ and vice-versa, given by $u^a\rightarrow iy^a$ and $y^b\rightarrow iu^b$ and corresponding to a signature inversion on only the transverse coordinates, interchanges the expressions for ${R^{u_a}}_{u_a}$ and ${R^{y_a}}_{y_a}$ and introduces a minus into all the curvature components, ${R^\mu}_\nu\rightarrow -{R^\mu}_\nu$. One could also achieve these signature changes by transforming the metric components: a signature inversion on all of the longitudinal coordinates $g^{t_it_i}\rightarrow -g^{x_ix_i}$ and $g^{x_ix_i}\rightarrow -g^{t_it_i}$ leaves the set of curvature terms unaltered, whereas $g^{u_au_a}\rightarrow -g^{y_ay_a}$ and $g^{y_ay_a}\rightarrow -g^{u_au_a}$ introduces a minus sign for all the curvature terms.

To find a new solution under longitudinal and transverse signature inversions the signs induced in the curvature components must match the sign changes in the remaining terms of the Einstein equations, those derived from the field strength and the dilaton. For the eleven dimensional case, which we consider in this paper, there is no dilaton, $\phi\rightarrow 0$, $a_i\rightarrow 0$, so we shall disregard it in the following discussion.

The field strength terms in each of the Einstein equations for the usual single brane solutions, with only one temporal coordinate longitudinal to the brane and none transverse, given in appendix \ref{electricbranesolutions}, are proportional to 
\begin{equation}
g^{tt'}g^{y_{1}y_{1}'}\ldots g^{y_{p}y_{p}'}g^{y_ay_a'}(F_{t'y_{1}'\ldots y_{p}'y_a'})(F_{ty_{1}\ldots y_{p}y_a})
\end{equation}
With the generalisation to our ansatz (\ref{ansatz}) to include multiple time coordinates longitudinal and transverse to the brane, the equivalent proportional term is 
\begin{equation}
g^{t_{1}t_{1}'}\ldots g^{t_{q}t_{q}'}g^{y_{1}y_{1}'}\ldots g^{y_{p}y_{p}'}g^{\mu \mu'}(F_{t_{1}'\ldots t_{q}'y_{1}'\ldots y_{p}'\mu'})(F_{t_{1}\ldots t_{q}y_{1}\ldots y_{p}\mu}) \label{rhs}
\end{equation}
Where the radial coordinate $\mu$ may now be a spatial or temporal transverse coordinate. An inversion of the longitudinal coordinates only, causes a sign change $(-1)^{p+q}$ in this term. The effect of inverting only the transverse coordinates $g^{\mu\mu'} \rightarrow -g^{\mu\mu'}$ introduces a minus sign as there is only one occurrence of the metric component with transverse coordinates in (\ref{rhs}).

A new solution is found under a signature inversion when the sign changes induced in the Riemann curvature components match the sign changes in the term (\ref{rhs}). For example, since we have observed that a signature inversion on only the transverse coordinates introduces a minus sign in both the curvature components and the remaining terms in the Einstein equations (\ref{rhs}), then a solution with signature components $[(q,p),(c,d)]$, will always find a new solution under an inversion of the transverse signature, taking the signature to $[(q,p),(d,c)]$. Additionally, if $p+q$ is even we may invert just the longitudinal coordinates and find another new solution $[(q,p),(c,d)]\rightarrow[(p,q),(c,d)]$, and furthermore in this case we may invert the full signature, an inversion of both longitudinal and transverse coordinates together,  $[(q,p),(c,d)]\rightarrow[(p,q),(c,d)]$ and find yet another new solution. However if $p+q$ is odd an inversion of the longitudinal signature introduces an unbalanced minus sign and no new solution is found.\footnote{However, we shall observe later that the $F^2$ term in the action may change its sign and in such cases the 'lost' solution of a $-F^2$ theory is a new solution of a $+F^2$ theory, but for the time being we continue to consider the usual $-F^2$ theory} For example, a solution in $(1,D-1)$ with longitudinal and transverse signature components $[(1,p),(0,D-p-1)]$ is the familiar $p$-brane solution. We find that for even $p+q$, we have the following set of signature components that carry a related solution, $[(1,p),(D-p-1,0)]$, $[(p,1),(0,D-p-1)]$ and $[(p,1),(D-p-1,0)]$ which give spacetime signatures $(D-p,p)$, $(p, D-p)$ and $(1,D-1)$ respectively. For the case of odd $(p+q)$ we have only one alternative signature, coming from an inversion of only the transverse coordinates, which gives a new solution, namely $(D-p,p)$ with longitudinal and transverse components $[(1,p),(D-p-1,0)]$.

Following the preceding considerations we are in a position to write down a set of signatures in which the $M2$ and $M5$ branes remain solutions. The $M2$ brane has $p+q=3$ and hence has related solutions in $(1,10)$ and $(9,2)$, with the following parameters,
\begin{center}
\begin{tabular}{c|c|c}
Signature & Longitudinal & Transverse \\
& Signature & Signature \\
\hline
$(1,10)$ & $(1,2)$ & $(0,8)$\\
$(9,2)$ & $(1,2)$ & $(8,0)$\\
\end{tabular}
\end{center}
Equivalently, the $M5$ brane has $p+q=6$ and has related solutions in $(1,10)$, $(6,5)$, $(5,6)$ and $(10,1)$,
\begin{center}
\begin{tabular}{c|c|c}
Signature & Longitudinal & Transverse \\
& Signature & Signature \\
\hline
$(1,10)$ & $(1,5)$ & $(0,5)$\\
$(6,5)$ & $(1,5)$ & $(5,0)$\\
$(5,6)$ & $(5,1)$ & $(0,5)$\\
$(10,1)$ & $(5,1)$ & $(5,0)$\\
\end{tabular}
\end{center}
In addition to inverting components of the signature we may also transform individual temporal coordinates into spacelike coordinates and vice versa. It is observed by following the computations in appendix \ref{electricbranesolutions} and the term in equation (\ref{rhs}), that a given solution will give a new solution by converting an even number of longitudinal temporal coordinates into longitudinal spatial coordinates, while leaving the transverse coordinates unaltered. Such a transformation introduces a sign change $(-1)^{2m}=1$ into term (\ref{rhs}), where $m$ is an integer such that $q\pm2m,p\mp2m\geq 0$, and only alters the counting symbols ${\hat{\delta}^{t_i}}_{\hspace{7pt} t_i}$ and ${\hat{\delta}^{x_i}}_{\hspace{7pt} x_i}$ in the curvature components so that a new solution is found. That is, given a solution in a signature with components $[(q,p),(c,d)]$ then functions carrying the signature $[(q\pm2m,p\mp2m),(c,d)]$ will give a new solution. Applying this to each of the signatures containing solutions related to the $M2$ brane, we find the additional solutions,
\begin{center}
\begin{tabular}{c|c|c}
Signature & Longitudinal & Transverse \\
& Signature & Signature \\
\hline
$(3,8)$ & $(3,0)$ & $(0,8)$\\
$(11,0)$ & $(3,0)$ & $(8,0)$\\
\end{tabular}
\end{center}
And similarly, additional signatures for the $M5$ solutions are,
\begin{center}
\begin{tabular}{c|c|c}
Signature & Longitudinal & Transverse \\
& Signature & Signature \\
\hline
$(3,8)$ & $(3,3)$ & $(0,5)$\\
$(8,3)$ & $(3,3)$ & $(5,0)$\\
\end{tabular}
\end{center}
We can carry this argument to the transverse coordinates, but we are no longer restricted to transforming even multiples of temporal coordinates into spatial coordinates, indeed any integer is possible giving a range of new solutions in signatures $[(q,p),(c\pm m, d \mp m)]$ where m is an integer such that $c\pm m, d \mp m \geq 0$. For the solutions related to the $M2$-brane we find further solutions,
\begin{center}
\begin{tabular}{c|c|c}
Signature & Longitudinal & Transverse \\
& Signature & Signature \\
\hline
$(2,9)$ & $(1,2)$ & $(1,7)$\\
$(3,8)$ & $(1,2)$ & $(2,6)$\\
$(4,7)$ & $(1,2)$ & $(3,5)$\\
$(5,6)$ & $(1,2)$ & $(4,4)$\\
$(6,5)$ & $(1,2)$ & $(5,3)$\\
$(7,4)$ & $(1,2)$ & $(6,2)$\\
$(8,3)$ & $(1,2)$ & $(7,1)$\\
$(4,7)$ & $(3,0)$ & $(1,7)$\\
$(5,6)$ & $(3,0)$ & $(2,6)$\\
$(6,5)$ & $(3,0)$ & $(3,5)$\\
$(7,4)$ & $(3,0)$ & $(4,4)$\\
$(8,3)$ & $(3,0)$ & $(5,3)$\\
$(9,2)$ & $(3,0)$ & $(6,2)$\\
$(10,1)$ & $(3,0)$ & $(7,1)$\\
\end{tabular}
\end{center}
And for the $M5$ solution we find the further related solutions,
\begin{center}
\begin{tabular}{c|c|c}
Signature & Longitudinal & Transverse \\
& Signature & Signature \\
\hline
$(2,9)$ & $(1,5)$ & $(1,4)$\\
$(3,8)$ & $(1,5)$ & $(2,3)$\\
$(4,7)$ & $(1,5)$ & $(3,2)$\\
$(5,6)$ & $(1,5)$ & $(4,1)$\\
$(4,7)$ & $(3,3)$ & $(1,4)$\\
$(5,6)$ & $(3,3)$ & $(2,3)$\\
$(6,5)$ & $(3,3)$ & $(3,2)$\\
$(7,4)$ & $(3,3)$ & $(4,1)$\\
$(5,6)$ & $(5,1)$ & $(1,4)$\\
$(7,4)$ & $(5,1)$ & $(2,3)$\\
$(8,3)$ & $(5,1)$ & $(3,2)$\\
$(9,2)$ & $(5,1)$ & $(4,1)$\\
\end{tabular}
\end{center}
This discussion is exhaustive, we have found all signatures that give solutions in $-F^2$ theories related to the $M2$ and $M5$-brane solutions of $M$-theory. We note that a universal shorthand for assessing whether or not a given signature contains a solution is to count the number of temporal longitudinal coordinates and if this is odd we have a solution to $-F^2$ theories. 

For the purposes of this paper it will be useful to consider not only the truncated action of the form given in equation (\ref{action}), but also theories constructed from an action under the double Wick rotation transforming $A_{\mu_1\ldots\mu_n}\rightarrow -iA_{\mu_1\ldots\mu_n}$. Such an action, upto Chern-Simons terms, looks identical to that of equation (\ref{action}) except that the sign of the kinetic $F^2$ term has changed from "$-$" to "$+$". Such a theory can be imagined as originating with an imaginary brane charge and will be relevant to our later consideration of spacelike branes. Let us now apply the reasoning of this section to such a $+F^2$ action. Adjusting the considerations of this section to $+F^2$ theories introduces an extra minus sign in front of the field strength terms in the Einstein equations (\ref{EinsteinEq}) and as a consequence into the term proportional to the non-curvature components given in equation (\ref{rhs}). In this case we will have a solution if the number of longitudinal time coordinates is even, meaning that exactly all the signatures not listed in this section, from the set of all signatures in 11-dimensions, will admit solutions to a $+F^2$ theory. 

Let us introduce a new term $\kappa$ to keep track of solutions in both $+F^2$ and $-F^2$ theories. For reasons that will become clear we use $f(\alpha_{11})$ to indicate the sign in front of the $F^2$ term, if $f(\alpha_{11})=0$ we have a $-F^2$ theory and if $f(\alpha_{11})=1$ we have a $+F^2$ theory. We define $\kappa$ to be
\begin{equation}
\kappa\equiv(-1)^{({\hat{\delta}^{t_i}}_{\hspace{7pt}t_i}+f(\alpha_{11}))} \label{kappa}
\end{equation}
If $\kappa=-1$ we have a brane solution, otherwise we do not; this criterion will be used to check for solutions throughout this paper. We note that this implies that there are no extremal $S$-branes (${\hat{\delta}^{t_i}}_{\hspace{7pt}t_i}=0$) in $-F^2$ theories ($f(\alpha_{11})=0$) indicating the known result \cite{BhattacharyaRoy,GutperleSabra} that $S$-branes in $M$-theory have an associated imaginary charge $||{\bf Q}||$ in equation (\ref{harmonicfunction}). More simply, the transformation $||{\bf Q}||\rightarrow i||{\bf Q}||$ induces the transformation $-F^2\rightarrow +F^2$.

\section{The Local Sub-Algebra and Weyl Reflections} \label{local_sub_algebra_and_weyl}
In the non-linear realisation of $E_{11}$ as a coset symmetry $E_{11}/H_{11}$, the local denominator sub-algebra, $H_{11}$, was chosen to be Cartan involution invariant, yielding the maximal compact sub-algebra. It was understood that a Wick rotation would then give a Lorentz invariant sub-algebra. We recall that the Cartan involution, $\Omega$, takes the generators of the positive roots, $E_i={K^i}_{i+1}$, to the negative of the generators of the negative roots, $-F_{i}=-{K^{i+1}}_i$ and vice-versa,
\begin{equation}
\Omega{(E_i)}=-F_i \qquad \text{and} \qquad \Omega{(F_i)}=-E_i
\end{equation}
such that the set of  generators, $E_i-F_i$, is invariant under the Cartan involution, $\Omega(E_i-F_i)=-F_i-(-E_i)=E_i-F_i$, and form a basis for the local denominator sub-algebra, $H_{11}$. It was noted \cite{EnglertHouart} that the Cartan involution can be generalised to what has been called the temporal involution, $\hat{\Omega}$, whose action is:
\begin{equation}
\hat{\Omega}({K^i}_{i+1})=-\epsilon_i{K^{i+1}}_i
\end{equation}
Where $\epsilon_i=\pm 1$. The generalisation redefines $H_{11}$ to be the sub-algebra left invariant under the temporal involution, as opposed to the Cartan involution; we denote the local sub-algebra invariant under the temporal involution as $\hat H_{11}$. This redefinition allows $\hat H_{11}$ to include non-compact generators, and in this way to differentiate between temporal and spatial coordinates, imposing a signature on the sub-algebra and a Lorentzian invariance. The information about which coordinates are timelike is carried by the new variable $\epsilon_i$. For example, we may impose a $(1,10)$ signature where $x^1$ is the temporal coordinate, by taking $\epsilon_1=-1$ and $\epsilon_i=1$ for the remaining spatial coordinates, giving:
\begin{equation}
\hat{\Omega}({K^1}_2)={K^2}_1, \quad \hat{\Omega}({K^i}_{i+1})=-{K^{i+1}}_i \qquad \text{For} \qquad 2\leq i \leq 10
\end{equation}
We find a basis for the local denominator algebra consisting of both compact and non-compact generators that is invariant under the temporal involution:
\begin{align}
\hat{\Omega}(E_1+F_1)&=-\epsilon_1F_1-\epsilon_1E_1=E_1+F_1 \\
\nonumber \hat{\Omega}(E_i-F_i)&=-\epsilon_iF_i+\epsilon_iE_i=E_i-F_i \qquad \text{For}\qquad 2\leq i \leq 10
\end{align}
It was recently pointed out by Keurentjes \cite{Keurentjes, Keurentjes2} that the temporal involution does not commute with the Weyl reflections, $S_i$, where,
\begin{equation}
S_i\beta \equiv \beta - 2\frac{(\alpha_i,\beta)}{(\alpha_i,\alpha_i)}\alpha_i
\end{equation}
Under the action of the Weyl group, the choice of local sub-algebra is not preserved and we obtain a new set of $\epsilon_i$'s, corresponding to a different temporal involution $\hat\Omega'$. This idea is described in detail elsewhere \cite{Keurentjes, Keurentjes2, EnglertHouart} and may be summarised as
\begin{align}
S_i\hat\Omega({K^j}_{j+1})=S_i(\epsilon_j{K^j}_{j+1})=\epsilon_j\rho_j{K^{j+1}}_j\equiv\hat\rho_j\Omega'({K^j}_{j+1})
\end{align}
Where $S_i{K^j}_{j+1}=\rho_j{K^{j+1}}_j$ and $\rho_j=\pm1$ arises because ${K^j}_{j+1}$ are representations of the Weyl group up to a sign. A consequence of the new temporal involution is that a new set of compact and non-compact generators form the basis of the local denominator sub-algebra $\hat H'_{11}$, potentially corresponding to a new set of temporal and spatial coordinates and signature. 

The Weyl reflections corresponding to nodes on the the gravity line of $E_{11}$ preserve the signature of spacetime while the reflection in the exceptional root $\alpha_{11}$ can change it. Keurentjes makes use of a ${\mathbb{Z}}_2$ valued function on the root lattice, $f$, which encodes the values of the $\epsilon_i$'s. We may regard the function $f$ as a member of the weight space, it may be written $f=\sum_i n_i\lambda_i$ where $\lambda_i$ is in the weight space. Its action on the simple roots is $(f,\alpha_i)=n_i$ where $n_i$ take the values $0$ or $1$, a value $f(\alpha_i)=1$ corresponds to a Chevalley generator ${K^i}_{i+1}$ which has $\epsilon_i=-1$. Put more simply, one of $x^i$ or $x^{i+1}$ is timelike and the other spacelike. Alternatively $f(\alpha_i)=0$ implies that the mixed coordinates, $x^i$ or $x^{i+1}$, are of the same type, either both timelike or both spacelike. It is worth highlighting that the root associated with the group element does not determine $f$, to obtain a putative solution we must specify both a position in the root lattice (a root) and a vector ($f$) encoding the signature; hitherto the group element has been used to find solutions in signature $(1,10)$.

Keurentjes offers a useful shorthand notation for following the effect of Weyl reflections upon the function $f$, which we will describe here before making some use of it. First the values of the function are written out on the Dynkin diagram, with the value $f(\alpha_i)$ written in the position of $\alpha_i$ on the diagram. As an example a $(1,10)$ signature might have the diagram:
\begin{align}
\nonumber &0\\
\nonumber 0\hspace{5pt} 0\hspace{5pt} 0\hspace{5pt} 1\hspace{5pt} 1\hspace{5pt} 0\hspace{5pt} 0\hspace{5pt} &0\hspace{5pt} 0\hspace{5pt} 0\\
\nonumber s\hspace{6pt} s\hspace{6pt} s\hspace{6pt} s\hspace{6pt} t\hspace{6pt} s\hspace{6pt} s\hspace{6pt} s&\hspace{6pt} s\hspace{6pt} s\hspace{6pt} s
\end{align}
We have indicated beneath the diagram with a series of $s$ (spatial) and $t$'s (temporal) the nature of the coordinates obtained by commencing with a spacelike $x^1$ on the far left of the gravity line.
But we may also consider the case where $x^1$ is a temporal coordinate, giving a mostly timelike set of coordinates. In general each signature diagram is ambiguous, representing both $(t,s,\pm)$ and $(s,t,\pm)$, but as we shall see these signatures do not always contain related solutions, so some care must be taken to specify the nature of one of the coordinates so that a signature diagram is not ambiguous. 

The ten values of $f$ on the gravity line gives the nature of all eleven dimensions of spacetime. The value of $f(\alpha_{11})$ plays no part in this, but it is argued in \cite{Keurentjes} that it determines the sign in front of the kinetic term in the action derived from the non-linear realisation. Specifically, in this paper, we impose the choice that $f(\alpha_{11})=0$ corresponds to the usual minus sign in front of $F^2$, while the alternative implies a plus sign in front of the $F^2$ term. It will be useful to follow the convention and label our signatures as $(t,s,\pm)$, where we will always write the number of timelike coordinates first and where $'+'$ implies that the sign of $F^2$ is negative, and $'-'$ that our action has a positive $F^2$ term. 

Let us find the effect of a Weyl reflection, $S_i$, on the function $f$, we have,
\begin{align}
\nonumber S_i(f)&=\sum_jn_j\lambda_j -\sum_jn_j(\alpha_i,\lambda_j)\alpha_i\\
&\equiv \sum_jm_j\lambda_j
\end{align}
Where $m_j$ are the components of $S_i(f)$. Taking the inner product with $\alpha_k$ gives the relation between $n_k$ and $m_k$
\begin{align}
m_k=n_k-n_iA_{ki}
\end{align}
Thus we find Keurentjes' diagrammatic prescription for following signature change: to apply $S_i$ to $f$ we simply add the value of $f(\alpha_i)$ to all the nodes it is connected to, and subsequently reduce modulo two. The reduction modulo two comes from the size of the fundamental lattice which has edge length $\left|-\alpha_i\right|+\left|+\alpha_i\right|=2\left|\alpha_i\right|$, so the sub-algebra repeats with this unit and we need only consider a version of it upto $\pm n2\left|\alpha_i\right|, n \in {\mathbb{Z}}$. For example in the signature diagram above, $S_i$ where $ i=\{ 1\ldots 3,6\ldots 11 \}$ have no effect upon the signature, whereas $S_4$ and $S_5$ bring about the following two signature diagrams respectively.
\begin{align}
\nonumber &0\\
\nonumber 0\hspace{5pt} 0\hspace{5pt} 1\hspace{5pt} 1\hspace{5pt} 0\hspace{5pt} 0\hspace{5pt} 0\hspace{5pt} &0\hspace{5pt} 0\hspace{5pt} 0\\
\nonumber s\hspace{6pt} s\hspace{6pt} s\hspace{6pt} t\hspace{6pt} s\hspace{6pt} s\hspace{6pt} s\hspace{6pt} s&\hspace{6pt} s\hspace{6pt} s\hspace{6pt} s\\
\nonumber &0\\
\nonumber 0\hspace{5pt} 0\hspace{5pt} 0\hspace{5pt} 0\hspace{5pt} 1\hspace{5pt} 1\hspace{5pt} 0\hspace{5pt} &0\hspace{5pt} 0\hspace{5pt} 0\\
\nonumber s\hspace{6pt} s\hspace{6pt} s\hspace{6pt} s\hspace{6pt} s\hspace{6pt} t\hspace{6pt} s\hspace{6pt} s&\hspace{6pt} s\hspace{6pt} s\hspace{6pt} s
\end{align}
\subsection{Invariant Roots and Signature Orbits}
Single brane solutions have been found from the decomposition of $E_{11}$ with respect to its gravity line, encoded within the group element (\ref{groupelement}). The prescription for finding the brane solution used so far \cite{West,CookWest} made use of the lowest weight generator in a representation. It was not clear how to interpret the other weights in a representation, in particular the highest weight. For the antisymmetric representations that we shall consider herein, other generators are reached from the lowest weight by a series of Weyl reflections, which in the light of the previous discussion implies a potential signature change. If we commence with the $R^{123}$ and $R^{123456}$ generators in $(1,10,+)$, where $x^1$ is the time coordinate, and raise them to their highest weights with the Weyl reflections $S^{-1}_0=(S_1)(S_2S_1)\ldots(S_{10}S_9\ldots S_1)$\footnote{$S_0=(S_{1}\ldots S_{10})\ldots(S_1S_2)(S_1)$ is the series of Weyl reflections that takes the highest weight to the lowest weight for all representations of $A_{10}$, so that $S_0R^{91011}=R^{123}$ and $S_0R^{67891011}=R^{123456}$.}, we find that the effect on the signature diagram is,
\begin{alignat}{2}
\nonumber &0& &1\\
 1\hspace{5pt} 0\hspace{5pt} 0\hspace{5pt} 0\hspace{5pt} 0\hspace{5pt} 0\hspace{5pt} 0\hspace{5pt} &0\hspace{5pt} 0\hspace{5pt} 0 &\qquad \stackrel{S^{-1}_0}{\longrightarrow} \qquad 0\hspace{5pt} 0\hspace{5pt} 0\hspace{5pt} 0\hspace{5pt} 0\hspace{5pt} 0\hspace{5pt} 0\hspace{5pt} &0\hspace{5pt} 0\hspace{5pt} 1 \label{signaturediagram}\\
\nonumber t\hspace{6pt} s\hspace{6pt} s\hspace{6pt} s\hspace{6pt} s\hspace{6pt} s\hspace{6pt} s\hspace{6pt} s&\hspace{6pt} s\hspace{6pt} s\hspace{6pt} s& t\hspace{7pt} t\hspace{7pt} t\hspace{7pt} t\hspace{7pt} t\hspace{7pt} t\hspace{7pt} t\hspace{7pt} t&\hspace{7pt} t\hspace{7pt} t\hspace{7pt} s
\end{alignat}
Up to insisting that $x^1$ is preserved as a time coordinate, we obtain the signature $(10,1,-)$, where the singled out spatial coordinate is in the longitudinal sector, specifically the spatial coordinate here is $x^{11}$, and the gauge fields in each case are $A_{91011}$ and $A_{67891011}$. From the observations of section \ref{solution_preserving}, it is known that the $M2$ brane has a related solution in $(10,1,-)$ with signature components $[(2,1),(8,0)]$, alternatively the $M5$ brane does not have a related solution $[(5,1),(5,0)]$ in $(10,1,-)$.

Some Weyl reflections may change the signature diagram, and even the signature, without changing the root, so there is an ambiguity about which signature the root and its associated solution exist in. We now consider the example of the exceptional root $\alpha_{11}$, associated with the generator $R^{91011}$, the highest weight of the `$M2$ representation' and find what we shall refer to as its signature orbit.

\subsubsection{Membrane Solution Signatures} \label{membranesolutionsignatures}

Let us first consider the trivial Weyl reflections of the gravity line on the root $\alpha_{11}$. It is noted that $\alpha_{11}$ is invariant under the reflections $\{S_1,\ldots S_7\}$ and $\{S_9,S_{10}\}$ and we may apply any number of these reflections, without altering the root, although we may trivially change the signature diagram but not the signature. Furthermore, we can observe from the signature diagram of the highest weight, shown on the right of (\ref{signaturediagram}), that as only $f(\alpha_{10})=1$ along the gravity line, only a series of reflections composed of $\{S_9,S_{10}\}$ may have an effect on the signature diagram without effecting the root. Explicitly, the only possible different signature diagrams that may be reached without changing the root are,
\begin{alignat}{2}
\nonumber &1& &1\\
\nonumber 0\hspace{5pt} 0\hspace{5pt} 0\hspace{5pt} 0\hspace{5pt} 0\hspace{5pt} 0\hspace{5pt} 0\hspace{5pt} &0\hspace{5pt} 0\hspace{5pt} 1 &\qquad \stackrel{S_{10}}{\longrightarrow} \qquad 0\hspace{5pt} 0\hspace{5pt} 0\hspace{5pt} 0\hspace{5pt} 0\hspace{5pt} 0\hspace{5pt} 0\hspace{5pt} &0\hspace{5pt} 1\hspace{5pt} 1 \\
\nonumber t\hspace{7pt} t\hspace{7pt} t\hspace{7pt} t\hspace{7pt} t\hspace{7pt} t\hspace{7pt} t\hspace{7pt} t&\hspace{7pt} t\hspace{7pt} t\hspace{7pt} s& t\hspace{7pt} t\hspace{7pt} t\hspace{7pt} t\hspace{7pt} t\hspace{7pt} t\hspace{7pt} t\hspace{7pt} t&\hspace{7pt} t\hspace{7pt} s\hspace{7pt} t\\
&1& &1\\
\nonumber 0\hspace{5pt} 0\hspace{5pt} 0\hspace{5pt} 0\hspace{5pt} 0\hspace{5pt} 0\hspace{5pt} 0\hspace{5pt} &0\hspace{5pt} 0\hspace{5pt} 1 &\qquad \stackrel{S_9S_{10}}{\longrightarrow} \qquad 0\hspace{5pt} 0\hspace{5pt} 0\hspace{5pt} 0\hspace{5pt} 0\hspace{5pt} 0\hspace{5pt} 0\hspace{5pt} &1\hspace{5pt} 1\hspace{5pt} 0 \\
\nonumber t\hspace{7pt} t\hspace{7pt} t\hspace{7pt} t\hspace{7pt} t\hspace{7pt} t\hspace{7pt} t\hspace{7pt} t&\hspace{7pt} t\hspace{7pt} t\hspace{7pt} s& t\hspace{7pt} t\hspace{7pt} t\hspace{7pt} t\hspace{7pt} t\hspace{7pt} t\hspace{7pt} t\hspace{7pt} t&\hspace{7pt} s\hspace{7pt} t\hspace{7pt} t
\end{alignat}
Here $x^1$ has been held as a temporal coordinate. The interpretation is that the trivial Weyl reflection in the roots of the gravity line shift the singled-out coordinate between $\{x^9,x^{10},x^{11}\}$, the longitudinal brane coordinates. While it is always true that the gravity line Weyl reflections do not alter the signature, it is not generally true that they do not alter the value of $f(\alpha_{11})$. For example, consider the root, $\alpha_8+\alpha_9+\alpha_{10}+\alpha_{11}$, associated to the generator $R^{8910}$, for which only a series of reflections composed of $\{S_8,S_9\}$ may alter the signature diagram without changing the root. In this example the reflection $S_8$ may change the value of $f(\alpha_{11})$ in addition to shifting the singled-out coordinate amongst the longitudinal coordinates. In general, the gravity line, or trivial, Weyl reflections preserve a signature $(t,s)$ but do not necessarily preserve whether we are working with a $-F^2$ or a $+F^2$ theory.

We now turn our attention to the non-trivial signature changes that may be applied to a root, $\beta$, without altering it and we outline here a prescription for finding alternative signatures without altering the root. In order to consider the effects of the reflection $S_{11}$ on the signature we first transform our root to a new root that is invariant under $S_{11}$, we call the series of Weyl reflections applied to acheive this $U$. Furthermore, we restrict ourselves to using only trivial Weyl reflections in this transformation, $U$, so that we only effect a non-trivial signature change after we have transformed to an $S_{11}$ invariant root. These restrictions identify a unique $S_{11}$ invariant root for a given non-zero coefficient of $\alpha_{11}$ in the simple root expansion of the root, $\beta$, or the level of $\beta$. At this stage $S_{11}$ may be applied without changing the root, but with the potential of altering the signature. Our original root in the new signature may be reobtained by applying $U^{-1}$. These steps allow an algorithmic exploration of the related signatures for a specific root. 

For $\alpha_{11}$, at level one, the $S_{11}$ invariant root is $\alpha_7+2\alpha_8+\alpha_9+\alpha_{11}$. It is obtained from $\alpha_{11}$ by reflections $S_8S_7S_9S_8\equiv U$, so that $\alpha_{11}=U^{-1}S_{11}U\alpha_{11}$, and a new class of signature diagrams is obtained that is not trivially related to the first class. This process is repeated for every trivially related signature diagram and in this way all possible Weyl reflections preserving $\alpha_{11}$ are applied and the associated set of signature diagrams including $(10,1,-)$ is obtained, we call this set the signature orbit of $\alpha_{11}$. An equivalent approach would be to apply all possible trivial reflections to $S_{11}U\alpha_{11}$, before transforming back to $\alpha_{11}$ with $U^{-1}$. This procedure is simply completed by a computer program, with the results shown in table \ref{sigorbitshighm2}, where we have only listed the cases where we have taken $x^1$ to be a temporal coordinate. The equivalent signature orbit where $x^1$ is taken to be spacelike is found by inverting all signatures, while keeping $f(\alpha_{11})$ constant. The signature orbits for the lowest weight, associated with generator $R^{123}$, are found by applying $(S_1\ldots S_{10})(S_1\ldots S_9)(S_1\ldots S_8)$ and the results are shown in table \ref{sigorbitslowm2}. 
\begin{center}
\begin{table}[htpb]
\begin{tabular}{c|c|c|c|c}
Signature & Longitudinal & Transverse & Trivially& \\
(temporal $x^1$)& Signature & Signature & Related & $\kappa$\\
&&& Signatures&\\
\hline
$(10,1,-)$ & $(2,1)$ & $(8,0)$ & $3$ & $-1$\\
$(9,2,-)$ & $(3,0)$ & $(6,2)$ & $21$ & $+1$\\
$(2,9,-)$ & $(0,3)$ & $(2,6)$ & $7$ & $-1$\\
$(6,5,-)$ & $(2,1)$ & $(4,4)$ & $105$ & $-1$\\
$(5,6,-)$ & $(1,2)$ & $(4,4)$ & $105$ & $+1$\\
$(6,5,-)$ & $(0,3)$ & $(6,2)$ & $21$ & $-1$\\
$(5,6,-)$ & $(3,0)$ & $(2,6)$ & $7$ & $+1$\\
$(9,2,-)$ & $(1,2)$ & $(8,0)$ & $3$ & $+1$\\
\hline
&&&$272$
\end{tabular}
\caption{The signature orbit of the root associated to $R^{91011}$}\label{sigorbitshighm2}
\end{table}

\end{center}
\begin{center}
\begin{table}[htpb]
\begin{tabular}{c|c|c|c|c}
Signature & Longitudinal & Transverse & Trivially& \\
(temporal $x^1$)& Signature & Signature & Related & $\kappa$\\
&&& Signatures&\\
\hline
$(1,10,+)$ & $(1,2)$ & $(0,8)$ & $1$ & $-1$\\
$(10,1,+)$ & $(2,1)$ & $(8,0)$ & $2$ & $+1$\\
$(9,2,+)$ & $(3,0)$ & $(6,2)$ & $15$ & $-1$\\
$(9,2,-)$ & $(3,0)$ & $(6,2)$ & $13$ & $+1$\\
$(6,5,+)$ & $(2,1)$ & $(4,4)$ & $70$ & $+1$\\
$(6,5,-)$ & $(2,1)$ & $(4,4)$ & $70$ & $-1$\\
$(5,6,+)$ & $(1,2)$ & $(4,4)$ & $35$ & $-1$\\
$(5,6,-)$ & $(1,2)$ & $(4,4)$ & $35$ & $+1$\\
$(5,6,+)$ & $(3,0)$ & $(2,6)$ & $13$ & $-1$\\
$(5,6,-)$ & $(3,0)$ & $(2,6)$ & $15$ & $+1$\\
$(2,9,-)$ & $(2,1)$ & $(0,8)$ & $2$ & $-1$\\
$(9,2,-)$ & $(1,2)$ & $(8,0)$ & $1$ & $+1$\\
\hline
&&&$272$
\end{tabular}
\caption{The signature orbit of the root associated to $R^{123}$} \label{sigorbitslowm2}
\end{table}
\end{center}
We have found the signature orbits of the highest and lowest weights in the membrane representation, but we have not checked whether each prescribed signature offers a solution to the Einstein and gauge equations. Using our observations of section \ref{solution_preserving} it is, in fact, a quick exercise to check all signatures and see if they offer a solution. We have evaluated $\kappa$, defined in equation (\ref{kappa}) for each putative solution given in the tables, and wherever we find $\kappa=-1$ we have a solution of the Einstein equations. If we count the number of trivially related signatures for these cases we notice that exactly half of the total signature orbit are solutions, that is $3+7+105+21=136$ solutions associated to the generator $R^{91011}$. For spacelike $x^1$ we find $21+105+7+3=136$ solutions too. For the lowest weight generator $R^{123}$ considered in table \ref{sigorbitslowm2}, we again find $1+15+70+35+13+2=136$ solutions for timelike $x^1$, and similarly $2+13+70+35+15+1=136$ solutions for spacelike $x^1$.

\subsubsection{Fivebrane Solution Signatures} \label{fivebranesolutionsignatures}
We now turn our attention to the representation that gives the $M5$ solution. Its highest weight generator is $R^{67891011}$ which is associated with the root $\beta\equiv\alpha_6+2\alpha_7+3\alpha_8+2\alpha_9+\alpha_{10}+2\alpha_{11}$. Only the Weyl reflections $\{S_5,S_{11}\}$ alter the root. The level two $S_{11}$-invariant root is obtained from $\beta$ by acting upon it with the series of reflections given by $U\equiv S_{10}S_9S_8S_7S_6S_5$ and the lowest weight representation, with generator $R^{123456}$, is obtained under the reflection $(S_1\ldots S_{10})\ldots(S_1\ldots S_5)$. The signature orbits are shown in tables \ref{highm5sigorbits} and \ref{lowm5sigorbits} respectively. Again the cases where $\kappa=-1$ give solutions, but we note that there is a difference to the $M2$ case when we consider the solutions for spacelike $x^1$. As before the spacelike $x^1$ case is found by a global signature inversion, however for the $M5$ case this does not bring about a change of sign in $\kappa$. The solutions for spacelike and timelike $x^1$ are no longer complementary but identical. For the highest weight representation we find $3+3+12+60+40+3+12+3=136$ solutions, the same number of solutions as for the $M2$ case, but for the lowest weight we find $5+1+3+15+40+40+15+3+5+1=128$ solutions.
\begin{center}
\begin{table}[htpb]
\begin{tabular}{c|c|c|c|c}
Signature & Longitudinal & Transverse & Trivially& \\
(temporal $x^1$)& Signature & Signature & Related & $\kappa$\\
&&& Signatures&\\
\hline
$(10,1,+)$ & $(5,1)$ & $(5,0)$ & $3$ & $-1$\\
$(10,1,-)$ & $(5,1)$ & $(5,0)$ & $3$ & $+1$\\
$(2,9,+)$ & $(1,5)$ & $(1,4)$ & $3$ & $-1$\\
$(2,9,-)$ & $(1,5)$ & $(1,4)$ & $3$ & $+1$\\
$(9,2,+)$ & $(5,1)$ & $(4,1)$ & $12$ & $-1$\\
$(9,2,-)$ & $(5,1)$ & $(4,1)$ & $12$ & $+1$\\
$(6,5,+)$ & $(3,3)$ & $(3,2)$ & $60$ & $-1$\\
$(6,5,-)$ & $(3,3)$ & $(3,2)$ & $60$ & $+1$\\
$(5,6,+)$ & $(3,3)$ & $(2,3)$ & $40$ & $-1$\\
$(5,6,-)$ & $(3,3)$ & $(2,3)$ & $40$ & $+1$\\
$(6,5,+)$ & $(5,1)$ & $(1,4)$ & $3$ & $-1$\\
$(6,5,-)$ & $(5,1)$ & $(1,4)$ & $3$ & $+1$\\
$(5,6,+)$ & $(1,5)$ & $(4,1)$ & $12$ & $-1$\\
$(5,6,-)$ & $(1,5)$ & $(4,1)$ & $12$ & $+1$\\
$(6,5,+)$ & $(1,5)$ & $(5,0)$ & $3$ & $-1$\\
$(6,5,-)$ & $(1,5)$ & $(5,0)$ & $3$ & $+1$\\
\hline
&&&$272$&
\end{tabular}
\caption{The signature orbit of the root associated to $R^{67891011}$} \label{highm5sigorbits}
\end{table}
\end{center}
\begin{center}
\begin{table}[htpb]
\begin{tabular}{c|c|c|c|c}
Signature & Longitudinal & Transverse & Trivially& \\
(temporal $x^1$)& Signature & Signature & Related & $\kappa$\\
&&& Signatures&\\
\hline
$(10,1,+)$ & $(5,1)$ & $(5,0)$ & $5$ & $-1$\\
$(1,10,+)$ & $(1,5)$ & $(0,5)$ & $1$ & $-1$\\
$(2,9,+)$ & $(1,5)$ & $(1,4)$ & $3$ & $-1$\\
$(2,9,-)$ & $(1,5)$ & $(1,4)$ & $2$ & $+1$\\
$(9,2,+)$ & $(5,1)$ & $(4,1)$ & $15$ & $-1$\\
$(9,2,-)$ & $(5,1)$ & $(4,1)$ & $10$ & $+1$\\
$(6,5,+)$ & $(3,3)$ & $(3,2)$ & $40$ & $-1$\\
$(6,5,-)$ & $(3,3)$ & $(3,2)$ & $60$ & $+1$\\
$(5,6,+)$ & $(3,3)$ & $(2,3)$ & $40$ & $-1$\\
$(5,6,-)$ & $(3,3)$ & $(2,3)$ & $60$ & $+1$\\
$(6,5,+)$ & $(5,1)$ & $(1,4)$ & $15$ & $-1$\\
$(6,5,-)$ & $(5,1)$ & $(1,4)$ & $10$ & $+1$\\
$(5,6,+)$ & $(1,5)$ & $(4,1)$ & $3$ & $-1$\\
$(5,6,-)$ & $(1,5)$ & $(4,1)$ & $2$ & $+1$\\
$(5,6,+)$ & $(5,1)$ & $(0,5)$ & $5$ & $-1$\\
$(6,5,+)$ & $(1,5)$ & $(5,0)$ & $1$ & $-1$\\
\hline
&&&$272$
\end{tabular}
\caption{The signature orbit of the root associated to $R^{123456}$} \label{lowm5sigorbits}
\end{table}
\end{center}
\subsubsection{$pp$-Wave Solution Signatures} \label{ppWave}
The $pp$-wave is treated in the same manner. Its highest weight is associated to the root $\beta=\alpha_1+\ldots \alpha_{10}$ and its lowest weight is associated to the root $S_0\beta=-\beta$. In both cases only the Weyl reflection $\{S_1,S_{10},S_{11}\}$ alter the root. However, the negative roots have generators of the form ${K^a}_b$, where $a>b$, which are projected out of the general group element of $E_{11}$, see equation (2.24) in \cite{West}, so the lowest weight representation has generator ${K^1}_2$, and associated root $\alpha_1$. We note that $\alpha_1$ is only altered by the Weyl reflections $\{S_1,S_2\}$, and is related to the highest weight by $\alpha_1=S_2S_3\ldots S_{10}\beta$. There are a number of possible level 0 $S_{11}$-invariant roots, we make use of $U\equiv S_8S_7S_6S_5S_4S_3S_2S_1$ to transform $\beta$ into $\alpha_9+\alpha_{10}$ and then effect the signature-changing Weyl reflection, $S_{11}$, before transforming back to $\beta$. The signature orbits containing the $M$-theory $pp$-wave coming from the root associated to the highest weight and the lowest weight with a positive root generator are listed in tables \ref{highppwave} and \ref{lowppwave} respectively. 

Our analysis of solutions to the Einstein equations from $\kappa$ is not appropriate for the $pp$-wave, instead we have a $pp$-wave solution if the ansatz given in appendix \ref{electricbranesolutions} is satisfied \cite{Argurio, Stelle}. From the tables we obtain $1+2+7=10$ solutions for the $pp$-wave for each weight of the representation and, in addition, we note that there is no associated $M'$-theory $pp$-wave, in our signature orbits.
\begin{center}
\begin{table}[htpb]
\begin{tabular}{c|c|c|c|c}
Signature & Longitudinal & Transverse & Signature & Trivially\\
(temporal $x^1$)& Signature & Signature & of $\Omega_9$ & Related \\
&&&& Signatures\\
\hline
$(10,1,-)$ & $(1,0)$ & $(0,1)$ & $(9,0)$ & $1$ \\
$(2,9,+)$ & $(1,0)$ & $(0,1)$ & $(1,8)$ & $2$ \\
$(2,9,-)$ & $(1,0)$ & $(0,1)$ & $(1,8)$ & $7$ \\
\hline
&&&&$10$
\end{tabular}
\caption{The signature orbit of the root associated to highest weight $pp$-wave} \label{highppwave}
\end{table}
\end{center}
\begin{center}
\begin{table}[htpb]
\begin{tabular}{c|c|c|c|c}
Signature & Longitudinal & Transverse & Signature & Trivially\\
(temporal $x^1$)& Signature & Signature & of $\Omega_9$ & Related \\
&&&& Signatures\\
\hline
$(1,10,+)$ & $(1,0)$ & $(0,1)$ & $(0,9)$ & $1$\\
$(9,2,-)$ & $(1,0)$ & $(0,1)$ & $(8,1)$ & $2$\\
$(9,2,+)$ & $(1,0)$ & $(0,1)$ & $(8,1)$ & $7$\\
\hline
&&&&$10$
\end{tabular}
\caption{The signature orbit of the root associated to lowest weight $pp$-wave} \label{lowppwave}
\end{table}
\end{center}
\subsection{$S$-Branes from a Choice of Local Sub-Algebra} \label{spacelikeinvolution}
Spacelike branes or $S$-branes were discovered as a constituent of string theory by Gutperle and Strominger \cite{GutperleStrominger}, who argued that they were a timelike kink in the tachyon field on the world volume of an unstable D-brane, or D-brane anti-D-brane pair. There is a wealth of literature on the rolling tachyon \cite{rollingtachyon} whose association with $S$-branes was first highlighted by Sen. Supergravity $S$-branes were found by Chen, Gal'tsov and Gutperle in arbitrary dimension, D, in \cite{ChenGaltsovGutperle} and in D=10 by Kruczenski, Myers and Peet in \cite{KruczenskiMyersPeet}. These solutions were shown to be equivalent under a coordinate transformation by Bhattacharya and Roy in \cite{BhattacharyaRoy}. General $S$-brane solutions in eleven dimensions were also found in reference \cite{Ohta2}, where intersection rules are also considered. We concentrate here on simply identifying the spacelike branes of $M$-theory and the related solutions in other theories that may be constructed from the brane spectrum of $E_{11}$. 

The group element (\ref{groupelement}) has been used to find brane solutions in exotic signatures by Weyl reflecting the known electric brane solutions of $M$-theory. The group element itself does not know which signature its associated solution exists in, indeed signature information comes from the choice of local sub-algebra. In our solutions we have singled out electric field strengths, those which always have a temporal coordinate, and used these as a starting point for the signature orbits of the previous section. It was observed in section \ref{membranesolutionsignatures} that the Weyl reflections that did not alter the root kept the temporal coordinate on the brane world-volume for the $M$-theory solutions, presenting an obstacle to finding $S$-branes from the electric solutions by Weyl reflecting the group element (\ref{groupelement}). Let's look at this in more detail. If one rotates the coordinates, using a Weyl reflection, to obtain a new root and associated gauge field, the effect of $S_i$ on the expansion of $\beta\cdot H$ in \ref{groupelement} is to interchange ${K^i}_i$ and ${K^{i+1}}_{i+1}$, where $i=1\ldots 10$, and in terms of the coordinate indices on the gauge potential the indices $x^i$ and $x^{i+1}$ are swapped. For example consider the lowest weight of the $M2$ representation with gauge field $A_{123}$, whose indices are transformed in the following manner,
\begin{align}
\nonumber A_{123}&\stackrel{S_1}{\longrightarrow}A_{123}\\
A_{123}&\stackrel{S_2}{\longrightarrow}A_{123}\\
\nonumber A_{123}&\stackrel{S_3}{\longrightarrow}A_{124}
\end{align}
If we pick $x^3$ to be the timelike coordinate we might think that to remove it from the gauge field would require a Weyl reflection in $S_3$. The effect of $S_3$ on the signature diagram is to change the timelike coordinate from  $x^3$ to $x^4$,
\begin{alignat}{2}
\nonumber &0& &0\\
\nonumber 0\hspace{5pt} 1\hspace{5pt} 1\hspace{5pt} 0\hspace{5pt} 0\hspace{5pt} 0\hspace{5pt} 0\hspace{5pt} &0\hspace{5pt} 0\hspace{5pt} 0 &\qquad \stackrel{S_{3}}{\longrightarrow} \qquad 0\hspace{5pt} 0\hspace{5pt} 1\hspace{5pt} 1\hspace{5pt} 0\hspace{5pt} 0\hspace{5pt} 0\hspace{5pt} &0\hspace{5pt} 0\hspace{5pt} 0 \\
\nonumber s\hspace{6pt} s\hspace{6pt} t\hspace{6pt} s\hspace{6pt} s\hspace{6pt} s\hspace{6pt} s\hspace{6pt} s&\hspace{6pt} s\hspace{6pt} s\hspace{6pt} s& s\hspace{6pt} s\hspace{6pt} s\hspace{6pt} t\hspace{6pt} s\hspace{6pt} s\hspace{6pt} s\hspace{6pt} s&\hspace{6pt} s\hspace{6pt} s\hspace{6pt} s
\end{alignat}
Consequently the new gauge field $A_{124}$ remains electric and similar considerations for each possible choice of timelike coordinate show that an electric gauge field remains electric under Weyl reflections. Importantly the $S$-brane solutions of $M$-theory are not related to the electric solutions by Weyl reflections, and are not found in the signature orbits of the usual electric solutions. 

However, given any real form of an algebra that leaves a Lorentzian form, e.g. $-t^2+x_1^2+\ldots x_{10}^2$, invariant we may consider the complex extension of the algebra such that a Euclidean form is left invariant. A specific example of how the generators transform is ${K^1}_2\rightarrow i{K^1}_2$ where $x^1$ is the temporal coordinate. To reintroduce a Lorentzian symmetry we apply the inverse transformation. For example to make $x^{10}$ temporal, we transform ${K^{10}}_{11}\rightarrow -i{K^{10}}_{11}$, and obtain a complexified version of the original set of generators preserving a real Lorentzian form, $t^2+x_1^2+\ldots +x_9^2-x_{10}^2$. The result is that the generators, $A_{123}, A_{123456}$ and ${K^1}_2$ used to find the electric solutions of appendix \ref{electricbranesolutions} become $iA_{123}, iA_{123456}$ and $i{K^1}_2$, as would be expected by the Wick rotations as in equation \ref{Wickrotation}, and all their indices are now spacelike. 

Equivalently, there is a different choice of local sub-algebra with a different set of generators, all real, that preserve the same Lorentzian form, that give an identical theory but with a different sign in front of $F^2$ in the action.  For example the $S2$ and $S5$-branes are solutions in signature $(1,10,-)$ with a real set of generators. Our Weyl reflections lead us to pick out the real form of the sub-algebra, and the sign of $F^2$. The $pp$-wave has a null field strength, hence we make use of the complex generators to find its spacelike solution. These spacelike solutions are verified in appendix \ref{spacelikebranesolutions}.

We may commence with the $S2$ and $S5$-brane solutions of $M$-theory given in appendix \ref{spacelikebranesolutions}, encoded in the group element and find their signature orbits. This solution is identical to commencing using a local subgroup, $H_{11}$, whose temporal coordinate with respect to our ansatz (\ref{ansatz}) is not part of the brane world-volume. The metric for the $M$-theory spacelike solution takes the same form as our ansatz (\ref{ansatz}), explicitly,
\begin{equation}
ds^2=A^2(\sum_{j=1}^{j=p}dx_{j}^2)+B^2(-du^2+\sum_{b=1}^{b=d}dy_{b}^2)
\end{equation}
Where $A$ and $B$, are as defined in (\ref{singlebranecoefficients}), using the harmonic function,
\begin{equation}
N_{(1,D-p-2)}=1+\frac{1}{D-p-3}\sqrt{\frac{\Delta}{2(D-2)}}\frac{\|\bf Q\|}{\hat{r}^{(D-p-3)}} 
\end{equation}
Where $\hat{r}^2=-(u^1)^2+(y^1)^2+\ldots (y^{D-p-2})^2$. For the $S2$ and $S5$ branes we have the associated field strengths
\begin{align}
F_{x_1x_2x_3\hat{r}}=&4\partial_{[\hat{r}}A_{x_1x_2x_3]}=\partial_{\hat{r}}A_{x_1x_2x_3}=\partial_{\hat{r}}{N_{(1,7)}^{-1}} \label{spacelikefieldstrengths}\\
\nonumber F_{x_1x_2x_3x_4x_5x_6\hat{r}}=&7\partial_{[\hat{r}}A_{x_1x_2x_3x_4x_5x_6]}=\partial_{\hat{r}}A_{x_1x_2x_3x_4x_5x_6}=\partial_{\hat{r}}{N_{(1,4)}^{-1}}
\end{align}
Weyl reflections of these solutions then give new orbits of possible solutions and completes the range of signature configurations related by $E_{11}$, in that we find solutions of $M$-theory where the temporal coordinate may be any of $\{x^1\ldots x^{11}\}$ for each weight. We list these signature orbits for the case of the highest and lowest weights of the $S2$-brane in the tables \ref{highs2}, \ref{lows2} and similarly for the $S5$-brane in tables \ref{highs5}, \ref{lows5}. 

The $pp$-wave solution distinguishes three sets of coordinates, namely a longitudinal coordinate, a transverse coordinate and the nine remaining coordinates in 11-dimensions, $\Omega_9$. Consequently there are two alternative choices of local sub-algebra that may be made, the first introduces a transverse time coordinate and the second introduces a time coordinate into $\Omega_9$, which we then relabel $\Omega_{(1,8)}$. The former case is similar to the original $pp$-wave solution under an interchange of $K\rightarrow -K$, having a line element:
\begin{equation}
ds^2=-(1+K){dt_1}^2+(1-K){dx_1}^2+2Kdt_1dx_1+d\Omega_9^2
\end{equation}
Consequently this second choice of local sub-algebra leads to the same solution as the usual sub-algebra, but the $pp$-wave in this case is completely out of phase with the original $pp$-wave. This solution is still dependent on a static harmonic function.

There is a time-dependent solution, which we label the $Spp$-wave, coming from the choice of local sub-algebra that introduces a time coordinate into $\Omega_9$, which we indicate by $\Omega_{(1,8)}$. The solution is given in appendix \ref{spacelikebranesolutions}. We list the signature orbit of the highest weight case in table \ref{hSppwave}, as in the case of the $pp$-wave the lowest weight signature orbit is identical.

Let us count the solutions in the $S2$-brane signature orbit from the M, $M*$ and $M'$-theories\footnote{That is, only those with signature $(1,10,\pm)$,$(10,1,\pm)$,$(2,9,\pm)$,$(9,2,\pm)$,$(5,6,\pm)$ and $(6,5,\pm)$}. From the highest weight signature orbit in table \ref{highs2}, we find $1+63+35+21=120$ solutions, where $x^1$ is timelike and $7+105+21+3=136$ where $x^1$ is spacelike. Recollect that we found $136$ solutions related to the equivalent highest weight $M2$ signature orbit, giving a total of $256$ solutions with timelike $x^1$ and $272$ with spacelike $x^1$; in all we have $528$ solutions from the root associated to the $R^{91011}$ generator. Similarly for the lowest weight associated to the $S2$-brane we find $3+50+31+6+5+25=120$ solutions with timelike $x^1$ and $5+62+25+10+3+31=136$ with spacelike $x^1$, giving a total of $528$ solutions from the root associated to the $R^{123}$ generator.

The $S5$-brane signature orbit coming from the highest weight given in table \ref{highs5} has $4+1+9+36+36+24+9=119$ solutions of the three $M$-theories for both timelike and spacelike $x^1$. If we include the $136$ solutions from the highest weight of the $M5$ representation for both timelike and spacelike $x^1$, we find a total of $510$ solutions associated to the $R^{67891011}$ generator. Perhaps most interesting are the results from the lowest weight generator $R^{123456}$; from the $S5$ signature orbit in table \ref{lows2} we find $2+30+60+20+10+6=128$ solutions to the three $M$-theories for both timelike and spacelike $x^1$. Recalling that we earlier counted $128$ solutions from the lowest weight signature orbit containing the $M5$ brane for each choice of $x^1$ giving a total of $512$ solutions associated to the $R^{123456}$ generator. We note that there is a difference between both the total number and type of solutions associated the generator $R^{a_1\ldots a_6}$ at different weights, in contrast to the $R^{a_1a_2a_3}$ generator to which a consistent set of solutions is associated at each weight.
\begin{center}
\begin{table}[htpb]
\begin{tabular}{c|c|c|c|c}
Signature & Longitudinal & Transverse & Trivially& \\
(temporal $x^1$)& Signature & Signature & Related & $\kappa$\\
&&& Signatures&\\
\hline
$(1,10,-)$ & $(0,3)$ & $(1,7)$ & $1$ & $-1$\\
$(10,1,-)$ & $(3,0)$ & $(7,1)$ & $7$ & $+1$\\
$(7,4,-)$ & $(2,1)$ & $(5,3)$ & $105$ & $-1$\\
$(4,7,-)$ & $(1,2)$ & $(3,5)$ & $63$ & $+1$\\
$(5,6,-)$ & $(2,1)$ & $(3,5)$ & $63$ & $-1$\\
$(6,5,-)$ & $(1,2)$ & $(5,3)$ & $105$ & $+1$\\
$(5,6,-)$ & $(0,3)$ & $(5,3)$ & $35$ & $-1$\\
$(6,5,-)$ & $(3,0)$ & $(3,5)$ & $21$ & $+1$\\
$(7,4,-)$ & $(0,3)$ & $(7,1)$ & $7$ & $-1$\\
$(4,7,-)$ & $(3,0)$ & $(1,7)$ & $1$ & $+1$\\
$(3,8,-)$ & $(0,3)$ & $(3,5)$ & $35$ & $-1$\\
$(8,3,-)$ & $(3,0)$ & $(5,3)$ & $21$ & $+1$\\
$(3,8,-)$ & $(2,1)$ & $(1,7)$ & $21$ & $-1$\\
$(8,3,-)$ & $(1,2)$ & $(7,1)$ & $3$ & $+1$\\
$(9,2,-)$ & $(2,1)$ & $(7,1)$ & $21$ & $-1$\\
$(2,9,-)$ & $(1,2)$ & $(1,7)$ & $3$ & $+1$\\
\hline
&&&$512$
\end{tabular}
\caption{The signature orbit of $S2$-brane from $E_{11}$ from $R^{91011}$} \label{highs2}
\end{table}
\end{center}
\begin{center}
\begin{table}[htpb]
\begin{tabular}{c|c|c|c|c}
Signature & Longitudinal & Transverse & Trivially& \\
(temporal $x^1$)& Signature & Signature & Related & $\kappa$\\
&&&Signatures&\\
\hline
$(10,1,+)$ & $(3,0)$ & $(7,1)$ & $3$ & $-1$\\
$(10,1,-)$ & $(3,0)$ & $(7,1)$ & $5$ & $+1$\\
$(7,4,+)$ & $(2,1)$ & $(5,3)$ & $50$ & $+1$\\
$(7,4,-)$ & $(2,1)$ & $(5,3)$ & $62$ & $-1$\\
$(4,7,+)$ & $(1,2)$ & $(3,5)$ & $25$ & $-1$\\
$(4,7,-)$ & $(1,2)$ & $(3,5)$ & $31$ & $+1$\\
$(5,6,+)$ & $(2,1)$ & $(3,5)$ & $62$ & $+1$\\
$(5,6,-)$ & $(2,1)$ & $(3,5)$ & $50$ & $-1$\\
$(6,5,+)$ & $(1,2)$ & $(5,3)$ & $31$ & $-1$\\
$(6,5,-)$ & $(1,2)$ & $(5,3)$ & $25$ & $+1$\\
$(8,3,+)$ & $(3,0)$ & $(5,3)$ & $31$ & $-1$\\
$(8,3,-)$ & $(3,0)$ & $(5,3)$ & $25$ & $+1$\\
$(4,7,+)$ & $(3,0)$ & $(1,7)$ & $5$ & $-1$\\
$(4,7,-)$ & $(3,0)$ & $(1,7)$ & $3$ & $+1$\\
$(9,2,+)$ & $(2,1)$ & $(7,1)$ & $10$ & $+1$\\
$(9,2,-)$ & $(2,1)$ & $(7,1)$ & $6$ & $-1$\\
$(2,9,+)$ & $(1,2)$ & $(1,7)$ & $5$ & $-1$\\
$(2,9,-)$ & $(1,2)$ & $(1,7)$ & $3$ & $+1$\\
$(6,5,+)$ & $(3,0)$ & $(3,5)$ & $25$ & $-1$\\
$(6,5,-)$ & $(3,0)$ & $(3,5)$ & $31$ & $+1$\\
$(3,8,+)$ & $(2,1)$ & $(1,7)$ & $6$ & $+1$\\
$(3,8,-)$ & $(2,1)$ & $(1,7)$ & $10$ & $-1$\\
$(8,3,+)$ & $(1,2)$ & $(7,1)$ & $3$ & $-1$\\
$(8,3,-)$ & $(1,2)$ & $(7,1)$ & $5$ & $+1$\\
\hline
&&&$512$
\end{tabular}
\caption{The signature orbit of $S2$-brane from $E_{11}$ from $R^{123}$} \label{lows2}
\end{table}
\end{center}
\begin{center}
\begin{table}[htpb]
\begin{tabular}{c|c|c|c|c}
Signature & Longitudinal & Transverse & Trivially& \\
(temporal $x^1$)& Signature & Signature & Related & $\kappa$\\
&&&Signatures&\\
\hline
$(1,10,-)$ & $(0,6)$ & $(1,4)$ & $4$ & $-1$\\
$(10,1,-)$ & $(6,0)$ & $(4,1)$ & $1$ & $-1$\\
$(7,4,+)$ & $(4,2)$ & $(3,2)$ & $36$ & $+1$\\
$(7,4,-)$ & $(4,2)$ & $(3,2)$ & $54$ & $-1$\\
$(4,7,+)$ & $(2,4)$ & $(2,3)$ & $24$ & $+1$\\
$(4,7,-)$ & $(2,4)$ & $(2,3)$ & $36$ & $-1$\\
$(5,6,+)$ & $(4,2)$ & $(1,4)$ & $6$ & $+1$\\
$(5,6,-)$ & $(4,2)$ & $(1,4)$ & $9$ & $-1$\\
$(6,5,+)$ & $(2,4)$ & $(4,1)$ & $24$ & $+1$\\
$(6,5,-)$ & $(2,4)$ & $(4,1)$ & $36$ & $-1$\\
$(5,6,+)$ & $(2,4)$ & $(3,2)$ & $54$ & $+1$\\
$(5,6,-)$ & $(2,4)$ & $(3,2)$ & $36$ & $-1$\\
$(6,5,+)$ & $(4,2)$ & $(2,3)$ & $36$ & $+1$\\
$(6,5,-)$ & $(4,2)$ & $(2,3)$ & $24$ & $-1$\\
$(7,4,+)$ & $(2,4)$ & $(5,0)$ & $9$ & $+1$\\
$(7,4,-)$ & $(2,4)$ & $(5,0)$ & $6$ & $-1$\\
$(3,8,-)$ & $(0,6)$ & $(3,2)$ & $6$ & $-1$\\
$(8,3,-)$ & $(6,0)$ & $(2,3)$ & $4$ & $-1$\\
$(3,8,+)$ & $(2,4)$ & $(1,4)$ & $9$ & $+1$\\
$(3,8,-)$ & $(2,4)$ & $(1,4)$ & $6$ & $-1$\\
$(8,3,+)$ & $(4,2)$ & $(4,1)$ & $36$ & $+1$\\
$(8,3,-)$ & $(4,2)$ & $(4,1)$ & $24$ & $-1$\\
$(9,2,+)$ & $(4,2)$ & $(5,0)$ & $6$ & $+1$\\
$(9,2,-)$ & $(4,2)$ & $(5,0)$ & $9$ & $-1$\\
$(2,9,+)$ & $(0,6)$ & $(2,3)$ & $4$ & $+1$\\
$(9,2,+)$ & $(6,0)$ & $(3,2)$ & $6$ & $+1$\\
$(4,7,+)$ & $(0,6)$ & $(4,1)$ & $4$ & $+1$\\
$(7,4,+)$ & $(6,0)$ & $(1,4)$ & $1$ & $+1$\\
\hline
&&&$510$&
\end{tabular}
\caption{The signature orbit of $S5$-brane from $E_{11}$ generator $R^{67891011}$} \label{highs5}
\end{table}
\end{center}
\begin{center}
\begin{table}[htpb]
\begin{tabular}{c|c|c|c|c}
Signature & Longitudinal & Transverse & Trivially& \\
(temporal $x^1$)& Signature & Signature & Related & $\kappa$\\
&&&Signatures&\\
\hline
$(10,1,+)$ & $(6,0)$ & $(4,1)$ & $3$ & $+1$\\
$(10,1,-)$ & $(6,0)$ & $(4,1)$ & $2$ & $-1$\\
$(7,4,+)$ & $(4,2)$ & $(3,2)$ & $40$ & $+1$\\
$(7,4,-)$ & $(4,2)$ & $(3,2)$ & $60$ & $-1$\\
$(4,7,+)$ & $(2,4)$ & $(2,3)$ & $20$ & $+1$\\
$(4,7,-)$ & $(2,4)$ & $(2,3)$ & $30$ & $-1$\\
$(5,6,+)$ & $(2,4)$ & $(3,2)$ & $20$ & $+1$\\
$(5,6,-)$ & $(2,4)$ & $(3,2)$ & $30$ & $-1$\\
$(6,5,+)$ & $(4,2)$ & $(2,3)$ & $40$ & $+1$\\
$(6,5,-)$ & $(4,2)$ & $(2,3)$ & $60$ & $-1$\\
$(3,8,+)$ & $(2,4)$ & $(1,4)$ & $15$ & $+1$\\
$(3,8,-)$ & $(2,4)$ & $(1,4)$ & $10$ & $-1$\\
$(8,3,+)$ & $(4,2)$ & $(4,1)$ & $30$ & $+1$\\
$(8,3,-)$ & $(4,2)$ & $(4,1)$ & $20$ & $-1$\\
$(5,6,+)$ & $(4,2)$ & $(1,4)$ & $30$ & $+1$\\
$(5,6,-)$ & $(4,2)$ & $(1,4)$ & $20$ & $-1$\\
$(6,5,+)$ & $(2,4)$ & $(4,1)$ & $15$ & $+1$\\
$(6,5,-)$ & $(2,4)$ & $(4,1)$ & $10$ & $-1$\\
$(7,4,+)$ & $(6,0)$ & $(1,4)$ & $3$ & $+1$\\
$(7,4,-)$ & $(6,0)$ & $(1,4)$ & $2$ & $-1$\\
$(8,3,+)$ & $(6,0)$ & $(2,3)$ & $4$ & $+1$\\
$(8,3,-)$ & $(6,0)$ & $(2,3)$ & $6$ & $-1$\\
$(9,2,+)$ & $(6,0)$ & $(3,2)$ & $4$ & $+1$\\
$(9,2,-)$ & $(6,0)$ & $(3,2)$ & $6$ & $-1$\\
$(2,9,+)$ & $(2,4)$ & $(0,5)$ & $5$ & $+1$\\
$(9,2,+)$ & $(4,2)$ & $(5,0)$ & $10$ & $+1$\\
$(4,7,+)$ & $(4,2)$ & $(0,5)$ & $10$ & $+1$\\
$(7,4,+)$ & $(2,4)$ & $(5,0)$ & $5$ & $+1$\\
\hline
&&&$510$
\end{tabular}
\caption{The signature orbit of $S5$-brane from $E_{11}$ generator $R^{123456}$} \label{lows5}
\end{table}
\end{center}
\begin{center}
\begin{table}[htpb]
\begin{tabular}{c|c|c|c|c|c}
Signature & Longitudinal & Transverse & Signature & Trivially& \\
(temporal $x^1$)& Signature & Signature & of $\Omega_9$ & Related & $\kappa$\\
&&&& Signatures&\\
\hline
$(10,1,-)$ & $(1,0)$ & $(1,0)$ & $(8,1)$ & $7$ & $+1$\\
$(10,1,+)$ & $(1,0)$ & $(1,0)$ & $(8,1)$ & $2$ & $-1$\\
$(4,7,-)$ & $(1,0)$ & $(1,0)$ & $(2,7)$ & $22$ & $+1$\\
$(4,7,+)$ & $(1,0)$ & $(1,0)$ & $(2,7)$ & $14$ & $-1$\\
\hline
&&&&$45$
\end{tabular}
\caption{The signature orbits of the root associated to highest weight $Spp$-wave} \label{hSppwave}
\end{table}
\end{center}
\section{Discussion}
It has previously been noted that the adjoint representation of $E_{11}$ may be decomposed to representations $A_{10}$ whose lowest weight has an associated group element (\ref{groupelement}) giving half BPS solutions to the usual $(1,10)$ theory. In this paper we considered representations of $E_{11}$ which were not the lowest weights under $A_{10}$. These can be related to the lowest weight by a series of $E_{11}$ Weyl reflections. The group element does indeed give solutions, but these are to theories which are not only in $(1,10)$, but also $(2,9)$, $(5,6)$ and their inverses, the transformation of signature being induced by Weyl reflections. We found that there exists an ambiguity in the signature for each solution due to there being a series of Weyl reflections that may preserve the root uniquely associated to the weight but may change the signature. We also carried out a general search for all the alternative signatures in which we found the equivalent of the $M2$, $M5$ and $pp$-wave solutions. The results of this search is probably not completely understood as some, and in most cases half, of the cases are not solutions, but would be under a change of sign of the $F^2$ term in the action for the theory, sign changes of $F^2$ being specified by the Weyl reflections \cite{Keurentjes}.

In particular the solutions we have found in signatures $(2,9)$ and $(5,6)$, derived from solutions of $M$-theory using the Weyl reflections of $E_{11}$, reproduce the solutions of $M*$ and $M'$ theories \cite{Hull}. The set of solutions to these theories bear an intimate relation to the string of weights of each representation, and in shifting between weights the number of solutions to each theory is, in general, not preserved. We considered the highest and lowest weight associated to the generators $R^{a_1a_2a_3}$ and $R^{a_1\ldots a_6}$ in sections \ref{membranesolutionsignatures} and \ref{fivebranesolutionsignatures} and counted the number of solutions in each case. It was noted that while the representation of the generator $R^{a_1a_2a_3}$ provided 136 solutions from both its highest and lowest weight, the representation of the generator $R^{a_1\ldots a_6}$ provided 136 solutions from its highest weight, but only 128 solutions from its lowest weight.

The solution counting differentiated between trivially related versions of the same solution, for example the $M2$-brane in $(1,10)$ with signature $[(1,2),(0,8)]$ contributes three related solutions to the count coming from the three $(={3\choose 1} \times {8\choose 0})$ different ways to associate a single timelike coordinate with the three coordinates $x^\mu$ of the brane world-volume. Of course in the signature $(1,10)$ there are eleven ways to choose the local sub-algebra, so one might expect to find eleven solutions in $(1,10)$ associated to each weight of the generator $R^{a_1a_2a_3}$ instead of three. The remaining eight solutions are the trivially related $S2$-brane solutions of signature $[(0,3),(1,7)]$. 

The arguments reviewed herein lead to the conclusion that readers who are predisposed to the $E_{11}$ conjecture are also encouraged to take up the $M*$ and $M'$-theories. Furthermore there appears to be no clear way to single out any of these three $M$-theories, as $E_{11}$ treats them all on an equal footing through the choice of local sub-algebra and the Weyl reflections.

In ordinary quantum field theory in Minkowski space one does require solutions in Euclidean space, namely instantons, which occur when the particle enters a region forbidden by energy considerations. One could imagine a similar scenario for branes and, in particular, one brane in the potential of another. We note that for every brane solution in the signature orbits given in this paper there is an exact pairing between putative solutions in $[(p,q),(c,d)]$ and $[(q,p),(c,d)]$, that is, under an signature inversion of the brane coordinates. For the fivebrane case either both or none of these signatures will carry solutions, while for the membrane case only one of the two signatures will carry a solution. Specifically the M5-brane, $[(1,5),(0,5)]=(1,10)$, is paired with a solution in $[(5,1),(0,5)]=(5,6)$ which is an $M'$-theory solution; and the M2-brane, $[(1,2),(0,8)]=(1,10)$, has a solution in $[(2,1),(0,8)]=(2,9)$, with a '$+F^2$' term, which is an $M*$-theory solution. From this viewpoint, we are always considering the theory in a $(1,10)$ signature but certain calulations would require theories in other signatures, namely $(2,9)$ and $(5,6)$.

{\centering \section*{Acknowledgements}}
PPC thanks Joe Gillard and Vid Stojevic for fruitful conversations. 

This research was supported by the PPARC grants PPA/S/S/2003/03644, PPA/G/O/2000/00451 and PPARC senior fellowship PPA/Y/S/2002/001/44.\\

\newpage
\appendix
\section{Electric Brane Solutions from $E_{11}$} \label{electricbranesolutions}
There is a substantial literature deriving single brane solutions in generic supergravity theories \cite{BPSsolutions}, for a review see \cite{Argurio, Stelle}. The $M2$, $M5$ and $pp$-wave solutions of $M$-theory have been derived from $E_{11}$ in \cite{West}.

For $E_{11}$ we have no dilaton and find the single brane solutions to be determined from a truncated action of the form
\begin{equation}
A=\frac{1}{16\pi G_{11}}\int {d^{11}x\sqrt{-g}(R-\frac{1}{2.4!}F_{a_1\ldots a_n}F^{a_1\ldots a_n})}
\end{equation}
$F_{a_1\ldots a_{n}}$ is a general $n$-form field strength formed in the non-linear realistion from the gauge fields. From \cite{West} we have two field strengths, which are dual to each other, which we consider as arising from the a 3-form and a 6-form gauge field
\begin{align}
F_{t_1x_1x_2r}=&4\partial_{[r}A_{t_1x_1x_2]}=\partial_rA_{t_1x_1x_2}=\partial_r{N_{(0,8)}^{-1}}\\
\nonumber F_{t_1x_1x_2x_3x_4x_5r}=&7\partial_{[r}A_{t_1x_1x_2x_3x_4x_5]}=\partial_rA_{t_1x_1x_2x_3x_4x_5}=\partial_r{N_{(0,5)}^{-1}}
\end{align}
The appropriate Einstein and gauge equations may be found by setting $\phi=0$, ${\hat{\delta}^{t_i}}_{\hspace{7pt} t_i}=1$, ${\hat{\delta}^{x_i}}_{\hspace{7pt} x_i}=p$, ${\hat{\delta}^{u_a}}_{\hspace{7pt} u_a}=0$ and ${\hat{\delta}^{y_a}}_{\hspace{7pt} y_a}=D-p-1$ in equations (\ref{EinsteinEq}) and (\ref{curvatureterms}).
The line element of an electric solution is derived from the solution generating group element and coincides with line element for single brane BPS solutions specified by equations (\ref{singlebranecoefficients}) \cite{West, CookWest}.
The non-linear realisation decomposes $E_{11}$ with respect to its longest gravity line $A_{10}$ obtaining a gravitational theory in 11 dimensions and an infinite array of irreducible representations. These representations are classified by level, the coefficient of the exceptional root $\alpha_{11}$ associated to the representation; all the usual solutions of eleven dimensional supergravity appear below level three. In the group element above, $\beta$ is the root associated to the lowest weight of each representation. The following solutions are found,
\subsection{The $M2$-Brane}
The line element of the $M2$-brane solution is \cite{DuffStelle}
\begin{equation}
ds^2=N^{-\frac{2}{3}}_{(0,8)}(-{dt_1}^2+dx_1^2+dx_2^2)+N^{\frac{1}{3}}_{(0,8)}(dy_1^2+\ldots dy_8^2) \label{$M2$metric}
\end{equation}
Giving a metric
\begin{displaymath}
{g_{\mu\nu}}=\bordermatrix{&t_1&x_1&x_2&y_1&\ldots &y_8\cr
t_1&-N_{(0,8)}^{-\frac{2}{3}}&0&0&0&0&0\cr
x_1&0&N_{(0,8)}^{-\frac{2}{3}}&0&0&0&0\cr
x_2&0&0&N_{(0,8)}^{-\frac{2}{3}}&0&0&0\cr
y_1&0&0&0&N_{(0,8)}^{\frac{1}{3}}&0&0\cr
\vdots&0&0&0&0&\ddots&0\cr
y_8&0&0&0&0&0&N_{(0,8)}^{\frac{1}{3}}\cr}
\end{displaymath}
So $\sqrt{-g}=N_{(0,8)}^{\frac{1}{3}}$ and we see that the gauge equation is satisfied by $F_{t_1x_1x_2y_a}=\partial_{y_a}{N_{(0,8)}^{-1}}$ in the following manner,
\begin{align}
\nonumber \partial_{y_a}(\sqrt{-g}F^{t_1x_1x_2y_a})=&\partial_{y_a}(N_{(0,8)}^{\frac{1}{3}}g^{t_1t_1'}g^{x_1x_1'}g^{x_2x_2'}g^{y_ay_a'}F_{t_1'x_1'x_2'y_a'})\\
\nonumber =&\partial_{y_1}(-N_{(0,8)}^2F_{t_1x_1x_2y_1})+\ldots \partial_{y_8}(-N_{(0,8)}^2F_{t_1x_1x_2y_8})\\
\nonumber =&\partial_{y_a}(-N_{(0,8)}^2\partial_{y_a}N_{(0,8)}^{-1}){\hat{\delta}^{y_a}}_{\hspace{7pt} y_a}\\
\nonumber =&\partial_{y_a}\partial_{y_a}N_{(0,8)}{\hat{\delta}^{y_a}}_{\hspace{7pt} y_a}\\
\nonumber =&0
\end{align}
In the last line we have used the fact that $N_{(0,8)}$ is an harmonic function in $y_a$, as can be checked from its definition in equation (\ref{harmonicfunction}). We see that the Einstein equations are satisfied by checking that the right-hand-side of each equation equals the curvature term for the electric solution. A term that appears frequently is $F^{t_1x_1x_2y_a}F_{t_1x_1x_2y_a}$ which we evaluate at the outset
\begin{align}
F^{t_1x_1x_2y_a}F_{t_1x_1x_2y_a}&=F^{t_1x_1x_2y_1}F_{t_1x_1x_2y_1}+\ldots F^{t_1x_1x_2y_8}F_{t_1x_1x_2y_8}\\
\nonumber &=-N_{(0,8)}^{-\frac{1}{3}}N_{(0,8)}^2(F_{t_1x_1x_2y_1})^2-\ldots -N_{(0,8)}^{-\frac{1}{3}}N_{(0,8)}^2(F_{t_1x_1x_2y_8})^2\\
\nonumber &=-8N_{(0,8)}^{-\frac{1}{3}}N_{(0,8)}^{-2}(\partial_{y_a}N_{(0,8)})^2
\end{align}
We proceed to check the Einstein equations,
\begin{align}
\nonumber \left(^{t_i}_{t_i}\right) \qquad \frac{1}{2.4!}(4&F^{t_1\mu_1\mu_2\mu_3}F_{t_1\mu_1\mu_2\mu_3}-\frac{1}{3}F^{\mu_1\mu_2\mu_3\mu_4}F_{\mu_1\mu_2\mu_3\mu_4})\\ 
&=\frac{1}{2.4!}(4.3!-\frac{4!}{3})F^{t_1x_1x_2y_a}F_{t_1x_1x_2y_a}\\
\nonumber &=-\frac{8}{3}N_{(0,8)}^{-\frac{1}{3}}N_{(0,8)}^{-2}(\partial_{y_a}N_{(0,8)})^2\\
\nonumber &=N^{-\frac{1}{3}}_{(0,8)} \{-\partial_{y_a}\partial_{y_a}\ln{N^{-\frac{1}{3}}_{(0,8)}} \}{\hat{\delta}^{y_a}}_{\hspace{7pt} y_a}\\
\nonumber &\equiv {R^{t_i}}_{t_i}\\
\nonumber \left(^{x_i}_{x_i}\right) \qquad 
\frac{1}{2.4!}(4&F^{x_i\mu_1\mu_2\mu_3}F_{x_i\mu_1\mu_2\mu_3}-\frac{1}{3}F^{\mu_1\mu_2\mu_3\mu_4}F_{\mu_1\mu_2\mu_3\mu_4}{\hat{\delta}^{x_i}}_{\hspace{7pt} x_i})\\
&=\frac{1}{2.4!}(4.3!-\frac{4!}{3})F^{t_1x_1x_2y_a}F_{t_1x_1x_2y_a}{\hat{\delta}^{x_i}}_{\hspace{7pt} x_i}\\
\nonumber &=N^{-\frac{1}{3}}_{(0,8)} \{-\partial_{y_a}\partial_{y_a}\ln{N^{-\frac{1}{3}}_{(0,8)}} \}{\hat{\delta}^{y_a}}_{\hspace{7pt} y_a}{\hat{\delta}^{x_i}}_{\hspace{7pt} x_i}\\
\nonumber &\equiv {R^{x_i}}_{x_i}\\
\nonumber \left(^{y_a}_{y_a}\right) \qquad 
\frac{1}{2.4!}(4&F^{y_a\mu_1\mu_2\mu_3}F_{y_a\mu_1\mu_2\mu_3}-\frac{1}{3}F^{\mu_1\mu_2\mu_3\mu_4}F_{\mu_1\mu_2\mu_3\mu_4})\\ 
&=\frac{1}{2.4!}(4.3!-\frac{4!}{3}{\hat{\delta}^{y_a}}_{\hspace{7pt} y_a})F^{t_1x_1x_2y_a}F_{t_1x_1x_2y_a}\\
\nonumber &=\frac{5}{6}N_{(0,8)}^{-\frac{1}{3}}N_{(0,8)}^{-2}{(\partial_{y_a}N_{(0,8)})}^2{\hat{\delta}^{y_a}}_{\hspace{7pt} y_a}\\
\nonumber &=N^{-\frac{1}{3}}_2 \{-8\partial_{y_a}\partial_{y_a}\ln{N^{\frac{1}{6}}_{(0,8)}}-3(\partial_{y_a}\ln{N^{-\frac{1}{3}}_{(0,8)}})^2\\
\nonumber & \qquad \qquad-6(\partial_{y_a}\ln{N^{\frac{1}{6}}_{(0,8)}})^2 \}{\hat{\delta}^{y_a}}_{\hspace{7pt} y_a}\\
\nonumber &\equiv {R^{y_a}}_{y_a}
\end{align}
\subsection{The $M5$-Brane}
The line element of the $M5$-brane solution is \cite{Guven}
\begin{equation}
ds^2=N^{-\frac{1}{3}}_{(0,5)}(-{dt_1}^2+dx_1^2+\ldots dx_5^2)+N^{\frac{2}{3}}_{(0,5)}(dy_1^2+\ldots dy_5^2)
\end{equation}
Giving a metric
\begin{displaymath}
{g_{\mu\nu}}=\bordermatrix{&t_1&x_1&\ldots &x_5&y_1&\ldots &y_5\cr
t_1&-N_{(0,5)}^{-\frac{1}{3}}&0&0&0&0&0&0\cr
x_1&0&N_{(0,5)}^{-\frac{1}{3}}&0&0&0&0&0\cr
\vdots&0&0&\ddots&0&0&0&0\cr
x_5&0&0&0&N_{(0,5)}^{-\frac{1}{3}}&0&0&0\cr
y_1&0&0&0&0&N_{(0,5)}^{\frac{2}{3}}&0&0\cr
\vdots&0&0&0&0&0&\ddots&0\cr
y_5&0&0&0&0&0&0&N_{(0,5)}^{\frac{2}{3}}\cr}
\end{displaymath}
So $\sqrt{-g}=N_{(0,5)}^{\frac{2}{3}}$ and we see that the gauge equation is satisfied by $F_{t_1x_1\ldots x_5y_a}=\partial_{y_a}{N_{(0,5)}^{-1}}$ in the following manner,
\begin{align}
\nonumber \partial_{y_a}(\sqrt{-g}F^{t_1x_1\ldots x_5y_a})=&\partial_{a}(-N_{(0,5)}^2\partial_aN_{(0,5)}^{-1}){\hat{\delta}^{y_a}}_{\hspace{7pt} y_a}\\
\nonumber =&\partial_{y_a}\partial_{y_a}N_{(0,5)}{\hat{\delta}^{y_a}}_{\hspace{7pt} y_a}\\
\nonumber =&0
\end{align}
As $N_{(0,5)}$ is an harmonic function in $y_a$. The Einstein equations are satisfied in the same way as the $M2$-brane solution, but it will be useful to express the equations in terms of the harmonic function $N_{(0,5)}$ for reference.
\begin{align}
\nonumber \left(^{t_i}_{t_i}\right) \hspace{7pt}\frac{1}{2.7!}(7.6!-\frac{2.7!}{3})F^{t_1x_1\ldots x_5y_a}F_{t_1x_1\ldots x_5y_a}=-\frac{1}{6}N_{(0,5)}^{-\frac{2}{3}}N_{(0,5)}^{-2}{(\partial_{y_a}N_{(0,5)})}^2{\hat{\delta}^{y_a}}_{\hspace{7pt} y_a}\\
\nonumber \left(^{x_i}_{x_i}\right) \hspace{7pt}\frac{1}{2.7!}(7.6!-\frac{2.7!}{3})5F^{t_1x_1\ldots x_5y_a}F_{t_1x_1\ldots x_5y_a}=-\frac{5}{6}N_{(0,5)}^{-\frac{2}{3}}N_{(0,5)}^{-2}{(\partial_{y_a}N_{(0,5)})}^2{\hat{\delta}^{y_a}}_{\hspace{7pt} y_a}\\
\nonumber \left(^{y_a}_{y_a}\right) \hspace{7pt}\frac{1}{2.7!}(7.6!-5\frac{2.7!}{3})F^{t_1x_1\ldots x_5y_a}F_{t_1x_1\ldots x_5y_a}=\frac{7}{6}N_{(0,5)}^{-\frac{2}{3}}N_{(0,5)}^{-2}{(\partial_{y_a}N_{(0,5)})}^2{\hat{\delta}^{y_a}}_{\hspace{7pt} y_a}
\end{align}
\subsection{The $pp$-Wave}
The $pp$-wave solution \cite{Hull2} arises from considering the lowest weight generator associated to a positive root, namely ${K^1}_2$, in the weight chain whose highest weight has root $\beta=\alpha_1+\alpha_2+\ldots \alpha_{10}$. ${K^1}_2$ is the generator of the root $\alpha_1$ and we find an associated line element, 
\begin{equation}
ds^2=-(1-K){dt_1}^2+(1+K){dx_2}^2-2Kdt_1dx_2+d\Omega_9^2
\end{equation}
Where we have made the substitution $N_{pp}=1+K$ and we note that $N_{pp}=1+\frac{\bf Q}{7r^7}$, and $r^2=y_1^2+\ldots y_9^2$. We note that $K=K(y_1,\ldots y_9)$, which is less general than the solution in \cite{Hull2}, but fits with the generic harmonic functions we have used for all brane solutions in this paper.

\section{Spacelike Brane Solutions from $E_{11}$} \label{spacelikebranesolutions}
In this appendix we demonstrate that the $S$-brane solutions discussed in section \ref{spacelikeinvolution} satisfy the Einstein equations (\ref{EinsteinEq}) in signature $(1,10,-)$ for our ansatz (\ref{ansatz}). As discussed in section  \ref{spacelikeinvolution} our field strength may be constructed out of a complexified version of the generators that give rise to the usual electric solutions in $(1,10,+)$, such that these solutions are derived from a truncated action with a $+F^2$ term. Equivalently we may use a real form of the sub-algebra generators in signature $(1,10,-)$ to construct our putative solutions. We follow the same approach as in appendix \ref{electricbranesolutions} and have two field strengths derived from a 3-form and a 6-form gauge field both of which have purely spatial indices, given in equation (\ref{spacelikefieldstrengths}). The appropriate Einstein and gauge equations may be found by setting $\phi=0$, ${\hat{\delta}^{t_i}}_{\hspace{7pt} t_i}=0$, ${\hat{\delta}^{x_i}}_{\hspace{7pt} x_i}=p+1$, ${\hat{\delta}^{u_a}}_{\hspace{7pt} u_a}=1$ and ${\hat{\delta}^{y_a}}_{\hspace{7pt} y_a}=D-p-2$ in equations (\ref{EinsteinEq}) and (\ref{curvatureterms}). The line element of a spacelike solution is derived from the solution generating group element in the same way as the electric case but using a choice of local sub-algebra that invokes a non-compact timelike generator in the transverse coordinates. The following solutions are associated with the lowest weights,
\subsection{The $S2$-Brane}
We now demonstrate that there exists an $S2$-brane solution in signature $(1,10,-)$ for our ansatz (\ref{ansatz}). The line element of the $S2$-brane solution is
\begin{equation}
ds^2=N^{-\frac{2}{3}}_{(1,7)}(dx_1^2+\ldots dx_3^2)+N^{\frac{1}{3}}_{(1,7)}(-du_1^2+dy_1^2+\ldots dy_7^2) \label{$S2$metric}
\end{equation}
Giving a metric
\begin{displaymath}
{g_{\mu\nu}}=\bordermatrix{&x_1&x_2&x_3&u_1&y_1&\ldots &y_7\cr
x_1&N_{(1,7)}^{-\frac{2}{3}}&0&0&0&0&0&0\cr
x_2&0&N_{(1,7)}^{-\frac{2}{3}}&0&0&0&0&0\cr
x_3&0&0&N_{(1,7)}^{-\frac{2}{3}}&0&0&0&0\cr
u_1&0&0&0&-N_{(1,7)}^{\frac{1}{3}}&0&0&0\cr
y_1&0&0&0&0&N_{(1,7)}^{\frac{1}{3}}&0&0\cr
\vdots&0&0&0&0&0&\ddots&0\cr
y_7&0&0&0&0&0&0&N_{(1,7)}^{\frac{1}{3}}\cr}
\end{displaymath}
So $\sqrt{-g}=N_{(1,7)}^{\frac{1}{3}}$ and we see that the gauge equation is satisfied by $F_{x_1x_2x_3\hat{r}}=\partial_{\hat{r}}{N_{(1,7)}^{-1}}$ in the following manner,
\begin{align}
\nonumber \partial_{\hat{r}}(\sqrt{-g}F^{x_1x_2x_3\hat{r}})=&\partial_{\hat{r}}(N_{(1,7)}^{\frac{1}{3}}g^{x_1x_1'}g^{x_2x_2'}g^{x_3x_3'}g^{\hat{r}\hat{r}'}F_{x_1'x_2'x_3'\hat{r}'})\\
\nonumber =&\partial_{u_1}(-N_{(1,7)}^2\partial_{u_1}N_{(1,7)}^{-1})+\partial_{y_a}(N_{(1,7)}^2\partial_{y_a}N_{(1,7)}^{-1}){\hat{\delta}^{y_a}}_{\hspace{7pt} y_a}\\
=&\partial_{u_1}\partial_{u_1}N_{(1,7)}-\partial_{y_a}\partial_{y_a}N_{(1,7)}{\hat{\delta}^{y_a}}_{\hspace{7pt} y_a} \label{S2gauge}\\
\nonumber =&0
\end{align}
In the last line we have used the fact that $N_{(1,7)}$ is an harmonic function in $y_a$, as can be checked from its definition in equation (\ref{harmonicfunction}),
\begin{align}
\nonumber -\partial_{u_1}\partial_{u_1}N_{(1,7)}+\partial_{y_a}\partial_{y_a}N_{(1,7)}{\hat{\delta}^{y_a}}_{\hspace{7pt} y_a}&=-\frac{6k}{\hat{r}^8}-\frac{8.6ku_1^2}{\hat{r}^{10}}-\frac{6k}{\hat{r}^8}{\hat{\delta}^{y_a}}_{\hspace{7pt} y_a}+\frac{8.6ky_a^2}{\hat{r}^{10}}{\hat{\delta}^{y_a}}_{\hspace{7pt} y_a}\\
&=-\frac{8.6k}{\hat{r}^8}+\frac{8.6k\hat{r}^2}{\hat{r}^{10}} \label{S2harmonic}\\
\nonumber &=0
\end{align}
Where $k=\pm\frac{\|\bf Q\|}{6}$ for the $S2$-brane, as defined in equation (\ref{harmonicfunction}).

We now check that the Einstein equations are satisfied by verifying that the right-hand-side of each equation equals the curvature term for the spacelike solution. A term that appears frequently is $F^{x_1x_2x_3\hat{r}}F_{x_1x_2x_3\hat{r}}$ which we evaluate at the outset
\begin{align}
\nonumber F^{x_1x_2x_3\hat{r}}F_{x_1x_2x_3\hat{r}}&=F^{x_1x_2x_3u_1}F_{x_1x_2x_3u_1}+F^{t_1x_1x_2y_1}F_{t_1x_1x_2y_1}+\ldots\\
&\qquad F^{t_1x_1x_2y_7}F_{t_1x_1x_2y_7}\\
\nonumber &=-N_{(1,7)}^{-\frac{1}{3}}N_{(1,7)}^2(F_{x_1x_2x_3u_1})^2+7N_{(1,7)}^{-\frac{1}{3}}N_{(1,7)}^2(F_{x_1x_2x_3y_a})^2
\end{align}
We proceed to check the Einstein equations,
\begin{align}
\nonumber \left(^{x_i}_{x_i}\right) \qquad 
-\frac{1}{2.4!}(4&F^{x_i\mu_1\mu_2\mu_3}F_{x_i\mu_1\mu_2\mu_3}-\frac{1}{3}F^{\mu_1\mu_2\mu_3\mu_4}F_{\mu_1\mu_2\mu_3\mu_4}{\hat{\delta}^{x_i}}_{\hspace{7pt} x_i})\\
&=-\frac{1}{2.4!}(4.3!-\frac{4!}{3})F^{x_1x_2x_3\hat{r}}F_{x_1x_2x_3\hat{r}}{\hat{\delta}^{x_i}}_{\hspace{7pt} x_i}\\
\nonumber &=N^{-\frac{1}{3}}_{(1,7)} \{\partial_{u_1}\partial_{u_1}\ln{N^{-\frac{1}{3}}_{(1,7)}}-\partial_{y_a}\partial_{y_a}\ln{N^{-\frac{1}{3}}_{(1,7)}}{\hat{\delta}^{y_a}}_{\hspace{7pt} y_a} \}{\hat{\delta}^{x_i}}_{\hspace{7pt} x_i}\\
\nonumber &\equiv {R^{x_i}}_{x_i}\\
\nonumber \left(^{u_a}_{u_a}\right) \qquad -\frac{1}{2.4!}(4&F^{u_1\mu_1\mu_2\mu_3}F_{u_1\mu_1\mu_2\mu_3}-\frac{1}{3}F^{\mu_1\mu_2\mu_3\mu_4}F_{\mu_1\mu_2\mu_3\mu_4})\\ 
&=-\frac{1}{2.4!}(4.3!F^{x_1x_2x_3u_1}F_{x_1x_2x_3u_1}-\frac{4!}{3}F^{x_1x_2x_3\hat{r}}F_{x_1x_2x_3\hat{r}})\\
\nonumber &=N_{(1,7)}^{-\frac{1}{3}}N_{(1,7)}^{-2}\{\frac{1}{3}(\partial_{u_1}N_{(1,7)})^2+\frac{7}{6}(\partial_{y_a}N_{(1,7)})^2\}\\
\nonumber &\equiv {R^{u_a}}_{u_a}\\
\nonumber \left(^{y_a}_{y_a}\right) \qquad 
-\frac{1}{2.4!}(4&F^{y_a\mu_1\mu_2\mu_3}F_{y_a\mu_1\mu_2\mu_3}-\frac{1}{3}F^{\mu_1\mu_2\mu_3\mu_4}F_{\mu_1\mu_2\mu_3\mu_4})\\ 
&=-\frac{1}{2.4!}(4.3!F^{x_1x_2x_3y_a}F_{x_1x_2x_3y_a}-\frac{4!}{3}{\hat{\delta}^{y_a}}_{\hspace{7pt} y_a}F^{x_1x_2x_3\hat{r}}F_{x_1x_2x_3\hat{r}})\\
\nonumber &=N_{(1,7)}^{-\frac{1}{3}}N_{(1,7)}^{-2}\{-\frac{1}{6}(\partial_{u_1}N_{(1,7)})^2+\frac{2}{3}(\partial_{y_a}N_{(1,7)})^2\}{\hat{\delta}^{y_a}}_{\hspace{7pt} y_a}\\
\nonumber &\equiv {R^{y_a}}_{y_a}
\end{align}

\subsection{The $S5$-Brane}
We now demonstrate that there exists an $S5$-brane solution in signature $(1,10,-)$ for our ansatz (\ref{ansatz}). The line element of the $S5$-brane solution is
\begin{equation}
ds^2=N^{-\frac{1}{3}}_{(1,4)}(dx_1^2+\ldots dx_6^2)+N^{\frac{2}{3}}_{(1,4)}(-du_1^2+dy_1^2+\ldots dy_4^2)
\end{equation}
Giving a metric
\begin{displaymath}
{g_{\mu\nu}}=\bordermatrix{&x_1&\ldots&x_6&u_1&y_1&\ldots &y_4\cr
x_1&N_{(1,4)}^{-\frac{1}{3}}&0&0&0&0&0&0\cr
\vdots&0&\ddots &0&0&0&0&0\cr
x_6&0&0&N_{(1,4)}^{-\frac{1}{3}}&0&0&0&0\cr
u_1&0&0&0&-N_{(1,4)}^{\frac{2}{3}}&0&0&0\cr
y_1&0&0&0&0&N_{(1,4)}^{\frac{2}{3}}&0&0\cr
\vdots&0&0&0&0&0&\ddots&0\cr
y_4&0&0&0&0&0&0&N_{(1,4)}^{\frac{2}{3}}\cr}
\end{displaymath}
So $\sqrt{-g}=N_{(1,4)}^{\frac{2}{3}}$ and we see that the gauge equation is satisfied by $F_{x_1\ldots x_6\hat{r}_a}=\partial_{\hat{r}_a}{N_{(1,4)}^{-1}}$ in the same manner as the field strength associated to the $S2$-brane in equations (\ref{S2gauge}) and (\ref{S2harmonic}). The Einstein equations are also satisfied in the same way as the $S2$-brane solution. We first note that
\begin{equation}
F^{x_1\ldots x_6\hat{r}}F_{x_1\ldots x_6\hat{r}}=-N_{(1,4)}^{-\frac{2}{3}}N_{(1,4)}^{-2}(\partial_{u_1}N_{(1,4)})^2+4N_{(1,4)}^{-\frac{2}{3}}N_{(1,4)}^{-2}(\partial_{y_a}N_{(1,4)})^2
\end{equation}
Let us now confirm that the Einstein equations in $(1,10,-)$ are satisfied for our ansatz (\ref{ansatz}).
\begin{align}
\nonumber \left(^{x_i}_{x_i}\right) \qquad 
-\frac{1}{2.7!}(7&F^{x_i\mu_1\ldots\mu_6}F_{x_i\mu_1\ldots\mu_6}-\frac{2}{3}F^{\mu_1\ldots\mu_7}F_{\mu_1\ldots\mu_7}{\hat{\delta}^{x_i}}_{\hspace{7pt} x_i})\\
&=-\frac{1}{2.7!}(7.6!-\frac{2.7!}{3})F^{x_1\ldots x_6\hat{r}}F_{x_1\ldots x_6\hat{r}}{\hat{\delta}^{x_i}}_{\hspace{7pt} x_i}\\
\nonumber &=N^{-\frac{2}{3}}_{(1,4)} \{\partial_{u_1}\partial_{u_1}\ln{N^{-\frac{1}{6}}_{(1,4)}}-\partial_{y_a}\partial_{y_a}\ln{N^{-\frac{1}{6}}_{(1,4)}}{\hat{\delta}^{y_a}}_{\hspace{7pt} y_a} \}{\hat{\delta}^{x_i}}_{\hspace{7pt} x_i}\\
\nonumber &\equiv {R^{x_i}}_{x_i}\\
\nonumber \left(^{u_a}_{u_a}\right) \qquad -\frac{1}{2.7!}(7&F^{u_1\mu_1\ldots\mu_6}F_{u_1\mu_1\ldots\mu_6}-\frac{2}{3}F^{\mu_1\ldots\mu_7}F_{\mu_1\ldots\mu_7})\\ 
&=-\frac{1}{2.7!}(7.6!F^{x_1\ldots x_6u_1}F_{x_1\ldots x_6u_1}-\frac{2.7!}{3}F^{x_1\ldots x_6\hat{r}}F_{x_1\ldots x_6\hat{r}})\\
\nonumber &=N_{(1,4)}^{-\frac{2}{3}}N_{(1,4)}^{-2}\{\frac{1}{6}(\partial_{u_1}N_{(1,4)})^2+\frac{4}{3}(\partial_{y_a}N_{(1,4)})^2\}\\
\nonumber &\equiv {R^{u_a}}_{u_a}\\
\nonumber \left(^{y_a}_{y_a}\right) \qquad 
-\frac{1}{2.7!}(7&F^{y_a\mu_1\ldots\mu_6}F_{y_a\mu_1\ldots\mu_6}-\frac{2}{3}F^{\mu_1\ldots\mu_7}F_{\mu_1\ldots\mu_7})\\ 
&=-\frac{1}{2.7!}(7.6!F^{x_1\ldots x_6y_a}F_{x_1\ldots x_6y_a}-\frac{2.7!}{3}{\hat{\delta}^{y_a}}_{\hspace{7pt} y_a}F^{x_1\ldots x_6\hat{r}}F_{x_1\ldots x_6\hat{r}})\\
\nonumber &=N_{(1,4)}^{-\frac{2}{3}}N_{(1,4)}^{-2}\{-\frac{1}{3}(\partial_{u_1}N_{(1,4)})^2+\frac{5}{6}(\partial_{y_a}N_{(1,4)})^2\}{\hat{\delta}^{y_a}}_{\hspace{7pt} y_a}\\
\nonumber &\equiv {R^{y_a}}_{y_a}
\end{align}
\subsection{The $Spp$-Wave}
The $Spp$-wave solution arises from considering the lowest weight in the weight chain whose highest weight has root $\beta=\alpha_1+\alpha_2+\ldots \alpha_{10}$, with a choice of local sub-algebra such that the temporal coordinate of $M$-theory is not one of the two distinguished coordinates as it is in the $pp$-wave solution. The line element derived from the group element (\ref{groupelement}) using $E_\beta=i{K^1}_2$ is a solution of the vacuum Einstein equations and is,
\begin{equation}
ds^2=(1-K)dx_1^2+(1+K)dx_2^2-2iKdx_1dx_2+d\Omega_{(1,8)}^2
\end{equation}
Where $d\Omega^2_{(1,8)}=-du_1^2+dy_1^2+\ldots dy_{8}^2$, and $K=K(u_1,y_1,\ldots y_8)$. This $Spp$-wave metric is the line element expected from a double Wick rotation of the $pp$-wave solution. A further Wick rotation would give a $pp$-wave solution of the $M*$-theory in $(2,9)$. It is non-static and has wavefronts that progress in the spacelike directions transverse to $\{x_1,x_2\}$.

\end{document}